\documentclass[aps,
prd,
english,
nofootinbib,
longbibliography,
showkeys,
reprint
]{revtex4-2}

\pdfoutput=1
\usepackage{amsfonts,amsmath,amssymb}
\usepackage{graphicx}
\usepackage[utf8]{inputenc}
\usepackage{hyperref}
\usepackage{babel}
\usepackage{listings}
\usepackage{matlab-prettifier}
\usepackage{enumitem}

\newcommand\e{{\rm e}}

\newcommand\be{\begin{equation}}
\newcommand\ee{\end{equation}}
\newcommand\bea{\begin{eqnarray}}
\newcommand\eea{\end{eqnarray}}

\begin{document}

\def\rhoo{\rho_{_0}\!} 
\def\rhooo{\rho_{_{0,0}}\!} 

\begin{flushright}
\phantom{
{\tt arXiv:2006.$\_\_\_\_$}
}
\end{flushright}

{\flushleft\vskip-1.4cm\vbox{\includegraphics[width=1.15in]{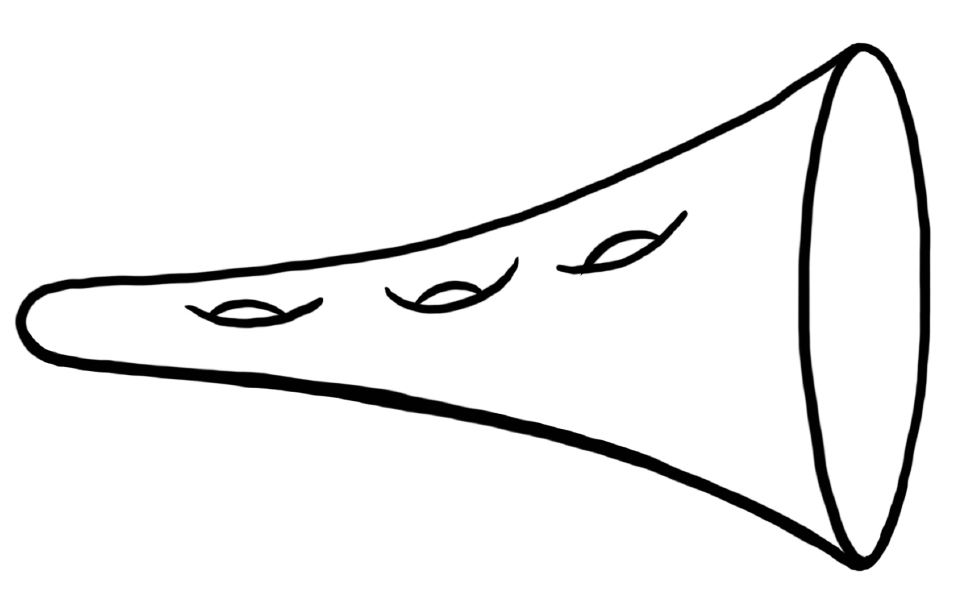}}}

\title{
  Fortuitous Chaos, BPS Black Holes, and  Random Matrices}
\author{Clifford V. Johnson}
\email{cliffordjohnson@ucsb.edu}

\affiliation{Department of Physics, Broida Hall,   University of California, 
Santa Barbara, CA 93106, U.S.A.}


\begin{abstract}
The ``fortuitous''  Bogomol'nyi-Prasad-Sommerfield (BPS)  sector states in gauge theory 
have been argued to furnish a description, through holography, of  generic BPS black hole microstates. They are expected to be strongly chaotic, a necessary feature to capture the black hole dynamics. This dovetails nicely with the existence of various random matrix models of JT supergravity with extended supersymmetry, within which the BPS chaos must be contained as a subsector. This paper studies a simple random matrix model that 
is a universal sector of all known random matrix models of JT supergravity with BPS sectors, arguing that it  captures many essential universal features of fortuitous BPS chaos. The model is  topological,  naturally interpolating between the Bessel and Airy models, where the gap energy $E_0$ controls the interpolation, and seems to have a simple  intersection theory interpretation.

\end{abstract}


\maketitle


\label{sec:introduction}
\section{Introduction}

This paper aims to strengthen  a triangle of relationships that has emerged
from recent key developments at the intersection of  black hole physics, large~${\rm \bar N}$  gauge theory, and quantum chaos and random matrix theory (RMT). The main focus will be on a {\it universal} model, offered as helpful in characterizing the chaos associated with certain Bogomol'nyi-Prasad-Sommerfield (BPS) sector states,  in a regime when $\Gamma$, the number of BPS states, is large. While the BPS states themselves live at zero energy, the energy spectrum of the model is peaked at just above the spectral gap known to be associated with the presence of the BPS states in gravity systems.   As  a random matrix model of positive $\Gamma{\times}\Gamma$ Hermitian matrices, the system is  manifestly strongly chaotic, and emerges as a special universal  sector of a variety of double-scaled matrix models known to describe JT supergravity with extended supersymmetry~\cite{Turiaci:2023jfa,Johnson:2023ofr,Johnson:2024tgg,Johnson:2025oty}. These in turn emerge at the throats of various supersymmetric black hole systems~\cite{Iliesiu:2020qvm,Heydeman:2020hhw,Heydeman:2025vcc}. The black holes in turn are holographically dual to so called ``fortuitous'' BPS sector states of a gauge theory~\cite{Chang:2024zqi}.
\begin{figure}
    \centering
\includegraphics[width=0.95\linewidth]{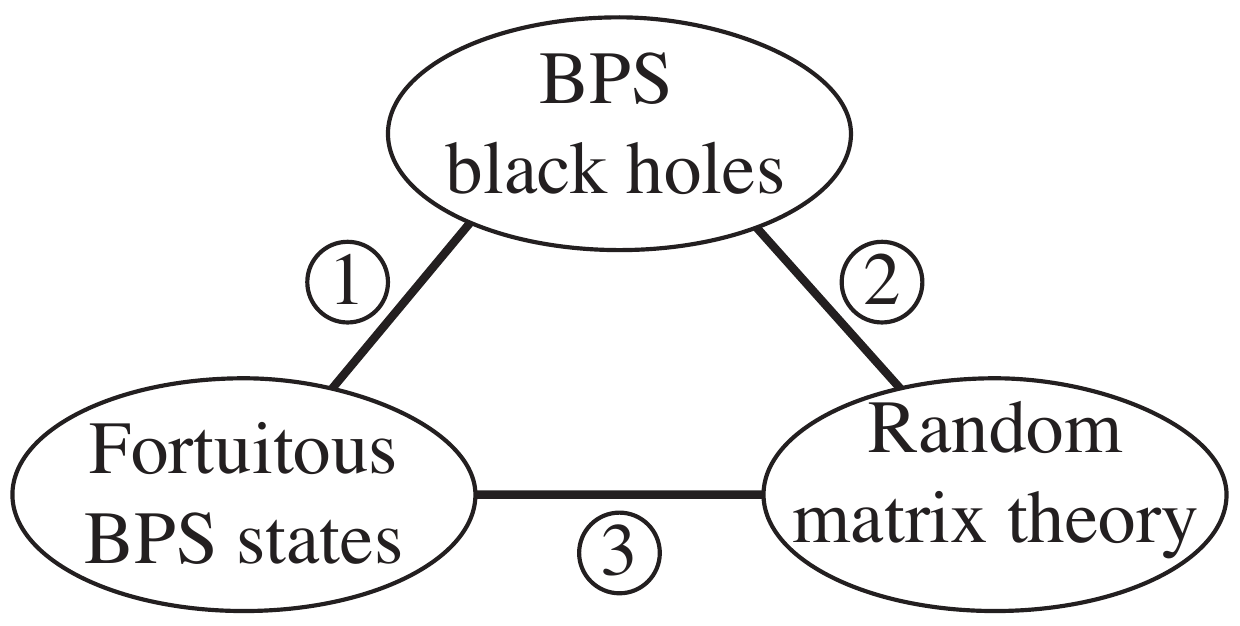}
    \caption{The triangle of connections.}
    \label{fig:triangle}
\end{figure}

These  relationships form a triangle,  shown in figure~\ref{fig:triangle}, with  topics at the vertices,and relationships as numbered sides. Sides~1 and~2   have been well-explored in the recent literature, and the point of this paper is that they imply, or perhaps demand, a {\it particular} RMT connection along side 3 that has not been hitherto highlighted  in the fortuity context. The model discussed here fills the gap.

$\bullet$ The left vertex  is occupied by  ``fortuitous'' BPS states~\cite{Chang:2024zqi} in gauge theory with extended supersymmetry, which  have become of keen interest recently. As ${\rm \bar N}$ (the rank of the gauge group) increases, such states can exit from the BPS spectrum since their existence relies on various trace relations being satisfied, and such relations simplify (and  disappear) at increasingly large~${\rm \bar N}$. There are also ``monotone'' BPS states that do not owe their existence to intricacies  of  trace relations at finite ${\rm \bar N}$. The  fortuitous states  are of interest here because the very nature of how they appear in the  BPS spectrum is expected to endow them with a  chaotic character. This is a precise notion of chaos, expected to be directly connected (relationship (1)) to the chaos associated with the BPS black holes that emerge in dual gravitational descriptions of the gauge theory at large~${\rm \bar N}$. (They are at the upper vertex of the triangle.) Quite broadly, and independently of the black hole connection, there should be a (model independent) way to characterize the chaos associated with these fortuitous states. There has been work on better understanding this ``BPS chaos'' in various systems, starting with {\it e.g.,} refs.~\cite{Lin:2022rzw,Lin:2022zxd,Chen:2024oqv,Chang:2024lxt}. In particular, the approach  of Lin, Maldacena, Rozenberg and Shan (LMRS)~\cite{Lin:2022rzw,Lin:2022zxd}  studies the matrix elements of  operators projected into the BPS part of the Hilbert space, and then examines whether they have random matrix statistics. While this can be done for the model discussed here (with interesting results), the main focus of this paper is to argue that signature features of its non-BPS energy spectrum are also  features of fortuity that arise in the gravitational limit. 

$\bullet$ At the upper vertex lives BPS black holes in quantum gravity. In the full theory (beyond just semiclassical considerations) such black holes are expected to enjoy a microscopic description of their degrees of freedom, and the holographically dual gauge theory is a natural place to find the appropriate microstates. Recent work has strongly suggested~\cite{Chang:2024zqi,Chen:2024oqv} that the gauge theory's fortuitous states are the generic microstates. Their chaotic spectrum is to be identified with the strongly chaotic spectrum black holes have somewhat generically\cite{Maldacena:2015waa}. By contrast, monotone states supply only a (non-chaotic, or at best, weakly chaotic) special subsector of the microstate description. Indeed, evidence has been presented\cite{Chen:2024oqv} that suggests that the  horizonless supergravity geometries of the fuzzball approach to black hole microstates\cite{Mathur:2024ify} correspond to monotone BPS states, or at best, states that are weakly chaotic. They are hence non-generic microstates.

The broad class of black holes in question possess a  finite area  horizon at extremality ($T{=} 0$) (yielding an extremal entropy, $S_0$), and the dynamics of the near-horizon geometry for low energy/temperature is governed by a (nearly) AdS$_2$ geometry. The  Schwarzian degree of freedom  appears at $T{\neq}0$, and can be thought of as appearing as a result of the spontaneous breaking of the $SL(2,\mathbb{R})$ symmetry of AdS$_2$. Its appearance~\cite{ Maldacena:2016hyu,Jensen:2016pah,Maldacena:2016upp,Engelsoy:2016xyb,Cotler:2016fpe} has striking consequences for the physics of near extremal black holes~\cite{Preskill:1991tb,Iliesiu:2020qvm,Heydeman:2020hhw}, as reviewed recently in ref.~\cite{Turiaci:2023wrh}. The overall description is that  Jackiw-Teitelboim (JT) gravity is the 2D low-energy description of the small $T$ throat dynamics. Crucially, the Euclidean gravitational path integral  description of JT gravity  has been shown to be equivalent  (perturbatively in $\hbar{=}\e^{-S_0}$)  to having a random matrix model (RMM) description~\cite{Saad:2019lba}, making manifest the chaotic nature of the low energy spectrum. This leads (relationship (2)) to the right vertex of the triangle.

$\bullet$   The JT/RMM correspondence has been sharpened in recent years, and includes new results in extended supergravity. BPS black holes have {\it e.g.}
${\cal N}{=}2$ and ${\cal N}{=}4$ JT supergravity in their throats~\cite{Iliesiu:2020qvm,Heydeman:2020hhw}, and, starting with the work of ref.~\cite{Turiaci:2023jfa}, extensive work has been done on the random matrix model description of such theories. In particular, ref.~\cite{Johnson:2023ofr} showed how to describe the ${\cal N}{=}2$ case in terms of multicritical potentials, with the (small) ${\cal N}{=}4$ given  similar treatment in ref.~\cite{Johnson:2024tgg}. In solving the matrix model string equation for that work, a core observation of ref.~\cite{Johnson:2023ofr}  was the class of ansatz for the solution needed to correctly describe  the BPS sector.  In fact, a miracle occurs in the calculation, showing that the sector modeling the BPS contribution works in concert with the non-BPS sector so as to give a consistent random matrix model {\it if and only if} the correct complete (non-BPS and BPS) JT supergravity spectrum is in play. This miracle was refined and distilled in ref.~\cite{Johnson:2025oty}, where it was applied to the new large ${\cal N}{=}4$ (and ${\cal N}{=}3$) models of ref.~\cite{Heydeman:2025vcc}.
In short, the connection between the BPS black holes and the random matrix models is clearly no accident, and it makes manifest their chaotic character.

 Given all that has been said above, clearly the next step is to establish the relationship (3) that constitutes the final side of the triangle. Here goes: {\it The aforementioned random matrix models of JT supergravity must contain within them a description of key features of fortuitous BPS states at large~${\rm \bar N}$.} 
In  particular one might expect that there ought to be a way of isolating from the black hole RMMs a stand-alone  model that characterizes key aspects of the ``fortuitous'' chaos of the BPS sector. The point of this paper is to highlight such a   model that naturally arises  by following this logic.\footnote{There is no claim here to be directly modelling the fortuity mechanism (of the gauge theory) itself. The random matrix models discussed in this paper do not seem complicated enough to be able to capture the needed ingredients (trace relations, {\it etc.}). Forthcoming work by another group~\cite{joaoEtAl} explores how matrix models can capture key features of the fortuity mechanism.} It is deceptively simple, but nevertheless non-trivial. What follows will  explore some of its features, and it will be proposed that, independently of the black hole context, the model contains a useful universal characterization of the chaos of BPS (fortuitous) states in the large ${\rm \bar N}$ limit.

It is worth  repeating one important point (not made clearly enough in the initial version of this paper). The focus of this paper is on the energy spectrum of the model, and even though the BPS states themselves are at zero energy, the idea is that the nature of the spectrum above zero, constituting the gap and  the peak that forms above it, are characteristic features of having a fortuitous BPS sector. The idea is that the matrix model describing these features  is a natural companion to the kinds of ``LMRS'' diagnostic random matrix models of BPS-projected operators~\cite{Lin:2022zxd,Lin:2022rzw}. While the latter are not the focus of the paper, we'll note later in Section~\ref{sec:LMRS-Jacobi} that there is a very natural choice of operator within the model for which the LMRS reduction produces an interesting class of random matrix models for which the eigenvalues are distributed according to the Jacobi ensemble~\cite{ForresterBook,Collins:2005} (also strongly chaotic). This is well worth further exploration.

\section{The Model: First Pass}
\begin{figure}[t]
    \centering
    \includegraphics[width=0.98\linewidth]{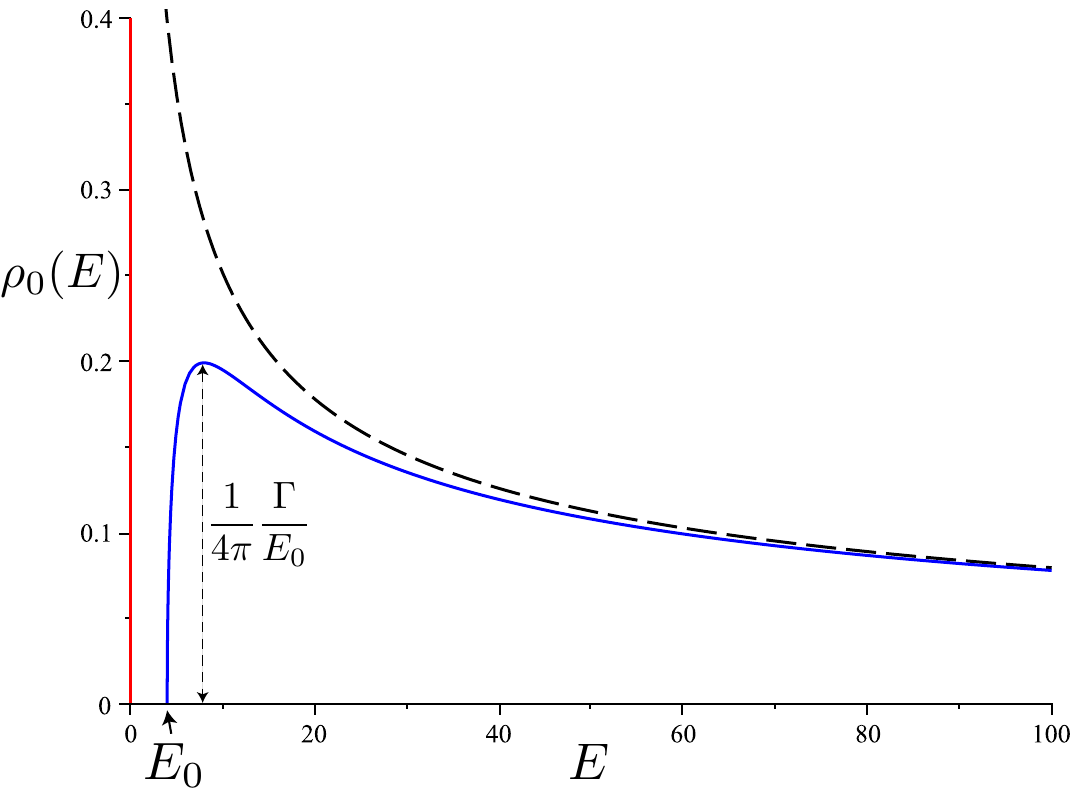}
    \caption{The leading density~(\ref{eq:leading}) showing (red line) the BPS states at $
    E{=}0$ and the accompanying continuum that starts after a gap of $E_0$ (equal to 4 here). There is a peak at $2E_0$ of height set by the ratio $\Gamma/E_0$. The dashed asymptote is the leading part of the ${\cal N}{=}1$ result $\rho_0(E){\sim} 1/\sqrt{E}$.}
    \label{fig:BPS-density}
\end{figure}

In this context, the model already made an appearance, at leading order, in refs.~\cite{Turiaci:2023jfa,Johnson:2023ofr}. In the approach of ref.~\cite{Johnson:2023ofr}, it explicitly formed the basis of the ansatz that showed how the ${\cal N}{=}2$ model can be constructed fully nonperturbatively in a  multicritical random matrix model approach. 
The leading   spectral density is:
\begin{equation}
    \label{eq:leading}
    \rho_0(E) = { \Gamma}\delta(E)+\frac{1}{2\pi\hbar}\frac{{\widetilde \Gamma}}{\sqrt{E_0}}\frac{\sqrt{E-E_0}}{E}\ .
\end{equation}
See figure~\ref{fig:BPS-density} for key features to be discussed in what follows.
 The  parameter $\Gamma$ is a measure of how many BPS states there are in the sector being described. They are at $E{=}0$. The combination ${\widetilde \Gamma}{=}\Gamma\hbar$ is natural to use as well, as will become more clear later.  The energy $E_0$  sets a gap above  $E{=}0$, after which a continuum of additional states appears.  It will be argued below that these states are special, come hand-in-hand with the BPS sector, and are an earmark of a fortuitous, chaotic BPS sector at least in the gravity limit. In a given model ${\widetilde\Gamma}$ and~$E_0$ can be functions of the R-charge of the sector in question,  the rank~${\rm \bar N}$ of the holographically dual gauge group, as well as the overall R-charge of the supercharge of the theory.\footnote{Consistency of the full matrix model description links them, as shown in ref.~\cite{Johnson:2024tgg}, and can be attributed to an extreme version of eigenvalue repulsion. In the supergravity context this means that the gap needed for consistency of the thermodynamics~\cite{Preskill:1991tb,Heydeman:2020hhw} is ensured by the RMM description.}
 The parameter $\hbar$ is a small parameter, scaling inversely with $N$, a measure of the size of the Hilbert space of the overall  system in which the physics is taking place. Typically, $\hbar\sim1/N$ is the (genus) expansion parameter of some double-scaled large $N$  random matrix model. (By the way, $N$ is not to be confused with the dual gauge theory's~${\rm \bar N}$). For example, in the black hole context, $\hbar{=}\e^{-S_0}{\sim}\frac{1}{N}$, as already recalled above. In terms of holographic dual parameters, $S_0{\sim}{{\rm \bar N}}^2$. Generally the $1/N$ parameter's role should also be  model independent. Whatever it is in a given model,  we work  in a limit where the number of BPS states is large (and where some  universality can therefore be hoped for). More precisely,  $\Gamma{=}\hbar^{-1}{\widetilde\Gamma}$, and so where large~$N$ matrix model  perturbation theory is reliable (small $\hbar$) we are describing a large number of BPS states. (Hence, for BPS black holes with a good supergravity description, the number of such BPS states is $O(\e^{S_0})$.)

The special continuum sector of the spectral density~(\ref{eq:leading}) deserves some attention, not the least because it seems superfluous to the BPS sector, and perhaps better left out of the discussion. In fact, a key point being made here is that it is {\it a crucial part} of the BPS story, and the chaotic nature of that sector.  This special continuum begins with a square root edge at $E_0$ (see figure~\ref{fig:BPS-density}), peaks at $E{=}2E_0$ at height $\frac{1}{4\pi}\frac{\Gamma}{E_0}$,  and subsequently  falls off with increasing energy, asymptoting to the classic $\frac{1}{\sqrt{E}}$ behaviour of Bessel models known to arise at the tail of  Wishart-type models with hard-edge behaviour. In fact,  the limit $E_0{\to}0$ with ${\widetilde\Gamma}{\to}0$ such that the ratio $\mu{=}{\widetilde\Gamma}/\sqrt{E_0}$ is held fixed, yields precisely the standard Bessel case everywhere. There are then  no BPS states and this is familiar from ${\cal N}{=}1$ models. (The convention is that $\mu{=}1$ there).

\section{The Model:  Second Pass}
A key observation to make next is that the  model is an effective statistical description of $\Gamma$ objects, and emerges as a decoupled  low energy limit of a larger class of  parent matrix models. For this to be tenable, the natural large parameter that should emerge is not the $N$ of the parent matrix model into which it can be embedded, but~$\Gamma$, the number of BPS states intrinsic to the low energy sector in question. This ought to be visible in how the random matrix perturbation theory organizes itself. A promising first sign is that the leading order piece in~(\ref{eq:leading}) has pre-factor of the form ${\tilde \hbar}^{-1}$, where:
\begin{equation}
    \label{eq:new-parameter}
    {\tilde\hbar}\equiv\frac{\hbar}{\mu}=\hbar\frac{\sqrt{E_0}}{\widetilde\Gamma}=\frac{\sqrt{E_0}}{\Gamma}\ ,
\end{equation}
{\it i.e.,} up to the factor $\sqrt{E_0}$, it looks like the expected disc contribution from a matrix model of rank $\Gamma$, with a  leading square root soft edge. We should look to higher orders to see if this organization persists. A natural method for swiftly computing  the corrections in the full matrix model's $\hbar$ perturbation theory has been developed recently~\cite{johnson:2024bue,Johnson:2024fkm,Lowenstein:2024gvz,Ahmed:2025lxe}, and is summarized in Appendix~\ref{app:correction}. Here are some of the results. Writing:
\begin{equation}
    \rho(E)=\frac{1}{\pi}{\rm Im} \sum_{g=0}^\infty R_{g}(E)+\text{non-perturb.,}
\end{equation} the method gives:
\begin{eqnarray}
\label{eq:R-terms}
    R_{0}(E) &=& \frac{1}{2{\tilde\hbar}}
    \frac{\sqrt{E_0-E}}{E}\ ,\nonumber\\
    R_{1}(E)  &=&- \frac{{\tilde\hbar}}{16}
    \frac{E}{(E_0-E)^{\frac52}}\ ,\nonumber\\
    R_{2}(E)  &=& -\frac{{\tilde\hbar}^3}{256}\frac{E(9E^2+80EE_0+16E_0^2)}{(E_0-E)^{\frac{11}{2}}}\ ,\nonumber\\
    R_{g}(E)  &=& -{{\tilde\hbar}^{2g-1}} \frac{E\times \text{Poly}_{2g-2}(E_0,E)}{(E_0-E)^{\frac{(1-6g)}{2}}}\ ,
\end{eqnarray}
where $\text{Poly}_{2g-2}(E_0,E)$ is an interesting polynomial in $E_0$ and $E$ that is of order $2g{-}2$. The first couple are listed above, and  the next is given by:
\begin{eqnarray}
\label{eq:extra-term}
 2048\times\text{Poly}_{4}(E_0,E)=&&\\
    &&\hskip-3.9cm 225 E^{4}+6160 E^{3} {E_0} +14160 E^{2} E_0^{2}+4352 E E_0^{3}+128 E_0^{4}\ .\nonumber
\end{eqnarray}
Happily, the $\hbar$ corrections all have just the right powers of $\widetilde\Gamma$ and $E_0$ in them to allow a reorganization into genus perturbation theory in $\tilde\hbar$. Why that happens will emerge naturally later, but for now the takeaway is that  an effective description of just  $\Gamma$ energies (the same number of elements of the BPS sector) has emerged in its own right, and just like other kinds of chaos with a RMM description, it has  a genus expansion!

Some other features are worth noting. First, imagine working on the complex $E$-plane, and write $E_0{-}E{=}z^2$, where  $z$  is the natural uniformizing parameter that appears due to the square root branch cut starting at~$E_0$. Normally, it transpires that $-2z R_{g}{=}W_{g,1}(z)$ where $W_{g,1}(z)$ is the one-boundary ($n{=}1$) subsector of the family of symplectic invariants $W_{g,n}(\{z_i\})$ ($i{=}1,\ldots,n$) associated with the spectral curve $({\bar x}(z),y(z))$ of the matrix model.\footnote{The variables $({\bar x},y)$ are used here instead of $(x,y)$ since $x$ will be used for a different variable in what is to come.}\cite{Eynard:2007kz,Eynard:2007fi,Eynard:2016yaa} The quantity $W_{g,n}$ comes with a factor $\hbar^{-\chi}{=}\hbar^{2g-2+n}$ where $\chi$ is the Euler number of the family of Riemann surfaces $\Sigma_{g,n}$ with~$g$ handles and $n$ marked points. $W_{g,n}$ has a natural topological ``intersection theory'' interpretation as the result of  integrating products of certain Chern classes that naturally are defined on the (compactified) moduli space ${\overline{\cal M}}_{g,n}$ of such surfaces. 

So what's happening in the model of interest here? The above corrections can be written suggestively in the following way: $-2z R_{g} {=}{\tilde \hbar}^{2g-1}{ W}_{g,1}(E_0,z)$, where $\tilde\hbar$ is our modified~$\hbar$ referring to size $\Gamma$ matrices instead of size $N$ (see~(\ref{eq:new-parameter}), and 
 the first few ${ W}_{g,1}(z)$ are:
\begin{eqnarray}
   \label{eq:Wgn-interpolator}
    &&\hskip-0.3cm {W}_{1,1}(z) = -\frac{1}{8 z^{2}}+\frac{E_0}{8 z^{4}}\ ,
\\
&&\hskip-0.3cm {W}_{2,1}(z) = -\frac{9 }{128 z^{4}}+\frac{107E_0}{128 z^{6}}-\frac{203 E_0^{2}}{128 z^{8}}+\frac{105 E_0^{3}}{128 z^{10}}\ ,
\nonumber\\
&&\hskip-0.3cm {W}_{3,1}(z) =-\frac{225}{1024 z^{6}}+\frac{7285 {E_0}}{1024 z^{8}}-\frac{20525 E_0^{2}}{512 z^{10}}+\frac{43021 E_0^{3}}{512 z^{12}}
\nonumber\\&&\hskip4.5cm-\frac{77077 E_0^{4}}{1024 z^{14}}+\frac{25025 E_0^{5}}{1024 z^{16}}\ .\nonumber
\end{eqnarray}
Reading off the highest order in $E_0$ in each 
case, we recognize the numbers (and the corresponding order in~$z$), as  the correlators $W_{g,1}^{\rm Airy}(z)$ of the classic Airy model, which has the spectral curve $({\bar x}{=}z^2/2, y {=} z)$ in topological recursion language.  In particular,
$W_{1,1}^{\rm Airy}(z){=}\frac18\frac{1}{z^4}, W_{2,1}^{\rm Airy}(z){=}\frac{105}{128}\frac{1}{z^{10}},$ and $  W_{3,1}^{\rm Airy}(z){=}\frac{25025}{1024}\frac{1}{z^{16}}$, {\it etc}.
On the other hand, zeroth order in $E_0$ has the familiar Bessel model  one-point correlators $W_{g,1}^{\rm Bessel}(z)$ defined by the spectral curve $({\bar x}{=}z^2/2, y {=} 1/z)$, {\it e.g.,} 
$W_{1,1}^{\rm Bessel}(z){=}{-}\frac18\frac{1}{z^2}, W_{2,1}^{\rm Bessel}(z){=}{-}\frac{9}{128}\frac{1}{z^4},$ and $  W_{3,1}^{\rm Bessel}(z){=}{-}\frac{225}{1024}\frac{1}{z^6}$.  

In between there is a series of terms with intermediate powers of $E_0$, which need an interpretation. To understand the origin of this structure, it is worth having a reminder of how the $W_{g,1}^{\rm Bessel/Airy}(z)$ are connected to intersection theory.\footnote{An excellent essay can be found in ref.~\cite{Do2008TouristGuide}, and ref.~\cite{Eynard:2016yaa} is a very useful book.}
Starting with Airy~\cite{Witten:1989ig,Witten:1990hr,Kontsevich:1992ti}, the key object is the first Chern class  of the  natural ``tautological'' bundle (one for each of the $n$ points; called $\psi$-classes, typically)  associated to $\overline{\cal M}_{g,n}$:
\begin{equation}
        \langle \tau_1\cdots\tau_{n}\rangle_g=\int_{\overline{\cal M}_{g,n}}\prod_{i=1}^n\psi_i^{d_i}\ ,
\end{equation}
and  $\sum_i d_i{=}{\rm dim}_\mathbb{C}\,\overline{\cal M}_{g,n}{=}3g{-}3{+}n$ for a non-zero integral, since a $\psi$-class has complex dimension 1. For one point the only non-zero integral is:
\begin{equation}
    \langle \tau_{3g-2}\rangle_g=\int_{\overline{\cal M}_{g,n}}\psi_1^{3g-2} = \frac{1}{(24)^gg!}\ ,
\end{equation} the classic result.
Meanwhile, the $W_{g,1}$ correlators package the  intersection numbers according to:
\begin{eqnarray}
    W_{g,1}^{\rm Airy}(z)dz&=& 
    (2d_1+1)!!\langle\tau_{3g-2} \rangle_g\frac{dz}{z^{2d_1+2}}
    \\&=&
    (6g-3)!!\langle\tau_{3g-2} \rangle_g\frac{dz}{z^{6g-2}}\ ,\nonumber
\end{eqnarray}
which gives the numbers seen above.
The Bessel model, by contrast, involves integrating $\psi$-classes against Norbury's $\Theta$-class, naturally defined for super-Riemann surfaces\cite{Norbury:2017eih}:
\begin{equation}
        \langle \tau_1\cdots\tau_{n}\cdot \Theta_{g,n}\rangle_g   
        =\int_{\overline{\cal M}_{g,n}}\Theta_{g,n}\prod_{i=1}^n\psi_i^{d_i}\ ,
\end{equation}
The object $\Theta_{g,n}$ turns out to have complex dimension $2g-2+n$ and so $\sum_id_i=g-1$. For $W_{g,1}$ we have:
\begin{equation}
    W_{g,1}^{\rm Bessel}(z)dz= -(2g-1)!!\langle\tau_{g-1} \Theta_{g,1}\rangle_g\frac{dz}{z^{2g}}\ ,
\end{equation}
and it transpires that\cite{Norbury:2017eih,guo2025combinatoricslargegenusasymptotics}:
 $   \langle\tau_{g-1} \Theta_{g,1}\rangle_g=\frac{((2g-1)!!)^2}{2^{3g}g!(2g-1)}\ ,
$  
reproducing the numbers seen above.\footnote{In the modern literature, these are often referred to as the Br\'{e}zin-Gross-Witten (BGW) numbers, computed by a certain unitary matrix model, sometimes with extra sources~\cite{Brezin:1980rk,Gross:1980he,Gross:1991aj}. This fits with the fact that it was established some time ago~\cite{Dalley:1992br} that such models are equivalent to certain random matrix models of Wishart form, the type naturally considered here (see next section) in a supersymmetric context.} 

Returning to understanding the structure of our results~(\ref{eq:Wgn-interpolator}), the organization suggests that our model of $\Gamma{\sim}\e^{S_0}$ BPS  states  has  a precise understanding in intersection theory as a  deformation of the Bessel model, with strength $E_0$. The first evidence is that    
the leading spectral density~(\ref{eq:leading})  defines (after the usual continuation of energy {\it via} $E{\to}{-}E$, and defining $z^2{=}E_0{-}E$) a problem with spectral curve $\Sigma$ given by:
\begin{equation}
\label{eq:new-spectral-curve}
 \Sigma:   \left({\bar x}=z^2{-}E_0\ ,y=\frac{z}{2(z^2-E_0)}\right)\ ,
\end{equation} which directly  interpolates between the Bessel ($E_0{=}0$ and Airy (large $E_0$) cases. 
The  terms in $W_{g,1}$ with intermediate powers of $E_0$  must  correspond to integrals over mixtures of additional powers of the pure $\psi$-classes with objects that replace $\Theta_{g,1}$. These objects  must have lower dimension than $\Theta_{g,1}$, such that non-zero integrals over~$\overline{\cal M}_{g,1}$ can result. In fact, if each power of $E_0$ counts such a replacement, it is easy to see that it must be replaced with something that has one fewer (complex) dimension, allowing the insertion of an extra~$\psi_1$, all the way up to removing all of $\Theta_{g,1}$'s $2g{-}1$  dimensions, allowing for $\psi_1^{3g-2}$ to saturate the integral, giving the Airy result with coefficient $E_0^{2g-1}$. This yields the expansion in $E_0$ seen at each genus in the examples in~(\ref{eq:Wgn-interpolator}). In summary, we ought to expect:
\begin{eqnarray}
  &&  W_{g,1}(z)dz= \\
  &&-\sum_{m=0}^{2g-1}(-E_0)^m(2g+2m-1)!!\langle\tau_{g+m-1} \Theta^{(m)}_{g,1}\rangle_g\frac{dz}{z^{2g+2m}}\ ,\nonumber
\end{eqnarray}
where we have denoted the expected new objects, with dimension $(2g{-}2+n{-}m)$ as $\Theta_{g,n}^{(m)}$, with $\Theta_{g,n}^{(0)}\equiv\Theta_{g,n}$.

This seems to amount to  a natural deformation of the $\Theta$-class itself. Happily  such matters have been studied in the mathematical literature and direct connection to the model of this paper has been made. (See {\it e.g.,}  sections 4.2 and 4.3 of ref.~\cite{Chidambaram:2022cqc} and the ``quantum Bessel'' model of section 2.3.5 in ref.~\cite{iwaki2018voroscoefficientshypergeometricdifferential}. See also ref.~\cite{bouchard2025thetaclassesgeneralizedtopological}.\footnote{Also relevant is the framework of the ``generalized Br\'{e}zin-Gross-Witten'' model discussed {\it e.g.} in refs.~\cite{Mironov:1994mv,Alexandrov:2016kjl,Dubrovin:2018cho,Alexandrov:2021etm}. Indeed, the generalization seems to be simply turning on $\widetilde\Gamma$ in the language of this paper, long known~\cite{Dalley:1992br} to be equivalent to adding extra quark flavors to the unitary matrix model.}) Presumably the ``descendant'' classes discussed in some of those works can play the role of the anticipated objects denoted $\Theta^{(m)}_{g,n}$ seen above. The numbers computed in~(\ref{eq:Wgn-interpolator}) then constitute (after stripping off the combinatorial factors) predictions for new intersection numbers involving them.  While  it is interesting to explore this further, such endeavours will be left   for future work.

Before leaving this, here is one more demonstration. Lest it be thought that this is all somehow special to the case for just one insertion ($n{=}1$) studied above, here are $W_{0,3}$, $W_{0,4}$ and $W_{1,2}$ for the model\footnote{After translating section 2.3.5 of ref.~\cite{iwaki2018voroscoefficientshypergeometricdifferential}, sending $z_i{\to} z_i/\sqrt{\lambda_0}$, and then $\lambda_0{\to}E_0$, and multiplying by $2$, it can be seen that their $W_{1,1}$ and $W_{2,1}$ match those in~(\ref{eq:Wgn-interpolator}) (computed here by the  method described in Appendix~\ref{app:correction}), and so their $W_{1,2}$ can be translated as well. But directly using topological recursion with the spectral curve~(\ref{eq:new-spectral-curve})  yields other examples. Note again that all $W_{g,1}$ can be swiftly obtained following Appendix~\ref{app:correction}'s methods, and all $W_{0,n}$ even more swiftly using  fact that for $n{\ge}3$ it essentially follows~\cite{Mertens:2020hbs} from $\partial_{x}^{n-3}u_0(x)$, after pulling off some trumpet factors. See section~5 of ref.~\cite{Ahmed:2025lxe}.}:
\begin{eqnarray}
\label{eq:W03-W12}
&&\hskip-0.35cm W_{0,3}(z_1,z_2,z_3)=
\frac{E_{0}}{z_{1}^{2} z_{2}^{2} z_{3}^{2}}\ ,\\
\nonumber
&&\hskip-0.35cm W_{0,4}(z_1,z_2,z_3,z_4) =-\frac{3 E_{0}}{z_{4}^{2} z_{3}^{2} z_{2}^{2} z_{1}^{2}}
    +\left(\sum^{4}_{i =1}\frac{1}{z_{i}^{2}}\right)\frac{3  E_{0}^{2}}{z_{4}^{4} z_{3}^{4} z_{2}^{4} z_{1}^{4}}\ ,\\
    &&\hskip-0.35cm W_{1,2}(z_1,z_2) =
    \frac{1}{32 z_{1}^{2} z_{2}^{2}}-\frac{\left(6 z_{1}^{4} z_{2}^{2}+6 z_{1}^{2} z_{2}^{4}\right) E_{0}}{32 z_{1}^{6} z_{2}^{6}}\nonumber
    \\
    &&\hskip3.5cm+\frac{\left(5 z_{1}^{4}+3 z_{1}^{2} z_{2}^{2}+5 z_{2}^{4}\right) E_{0}^{2}}{32 z_{1}^{6} z_{2}^{6}}\ , \nonumber
\end{eqnarray}
where again in each case, the Airy model correlator shows up at highest order in $E_0$, while that for Bessel is at zeroth order ($W_{0,3}{=}0$ for Bessel). The same structure was also verified for $W_{2,2}$, but it is too cumbersome to list here. Finally, the factor of 3 in $W_{0,4}$  is presumably a combinatorial factor coming from the three ways of choosing pairings out of $(z_1,z_2,z_3,z_4)$. In Feynman diagram language, $W_{0,3}$ is like a vertex that can be used to build higher $W_{0,n}$, and for $n{=}4$, two of them must be used, glued along an $s$, $t$ or $u$ channel.

Our exploration of the structure seen  to emerge from our  model's  intersection theory correlation functions was actually a fruitful  digression. It lies  at the heart of (and hence explains) something that had  been noticed for the analogous quantities in ${\cal N}{=}2$ JT supergravity, starting with 
Turiaci and Witten~\cite{Turiaci:2023jfa}, explored further (and extended to large and small ${\cal N}{=}4$ JT supergravity) in ref.~\cite{Ahmed:2025lxe}. After an inverse Laplace transform the JT supergravity $W_{g,n}$ become Weil-Petersson volumes $V_{g,n}$ (generalized for supergravity).  In   extended JT supergravity with gap~$E_0$, they also have an expansion in powers of~$E_0$. For example, for ${\cal N}{=}2$ the Weil-Petersson volume $V_{1,1}(b)$ was computed in ref.~\cite{Turiaci:2023jfa} as:
\begin{equation}
\label{eq:V11}
    V_{1,1}^{ {\rm JT}\,({\cal N}{=}2)}(b)=-\frac{1}{8}+\frac{(b^2+4\pi^2)}{48}E_0\ ,
\end{equation}
and $V_{2,1}(b)$ was computed in ref.~\cite{Ahmed:2025lxe} to be:
\begin{eqnarray}
\label{eq:V21}
V_{2,1}^{ {\rm JT}\,({\cal N}{=}2)}(b)=  -\frac{3\left(b^2 + 4\pi^2\right)}{256}&&
\\&&\hskip-4.9cm  + \frac{\left(b^2 + 4\pi^2\right)}{15360}  \left(107b^2 + 1772\pi^2\right)E_0 \nonumber \\
&&\hskip-4.9cm - \frac{\left(b^2 + 4\pi^2\right)}{92160}  \left(29b^4 + 1320b^2\pi^2 + 12592\pi^4\right)E_0^2\nonumber \\
&&\hskip-4.9cm + \frac{\left(b^2 + 4\pi^2\right)}{2211840}  \left(b^2 + 12\pi^2\right)\left(5b^4 + 384 b^2\pi^2 + 6960\pi^4\right)E_0^3\ .\nonumber
\end{eqnarray}

In these two  cases of  $V_{g,1}^{ {\rm JT}\,({\cal N}{=}2)}(b)$, ($g{=}1,2$), the coefficient of the highest power of $E_0$ can be recognized as the ordinary (bosonic) JT's  $V_{g,1}^{\rm JT}$, while the   $E_0$-independent part is  the $V_{g,1}^{ {\rm JT}\,({\cal N}{=}1)}$. 

For comparison, inverse Laplace transforming our special model's
$W_{1,1}(z)$ and $W_{2,1}(z)$ from (\ref{eq:Wgn-interpolator}) gives:
\begin{eqnarray}
\label{eq:little-volumes}\frac{1}{b}{\cal L}^{-1}(W_{1,1}(z),z,b)&=&-\frac{1}{8}+\frac{b^{2}}{48}E_{0} \,\\
    \frac{1}{b}{\cal L}^{-1}(W_{2,1}(z),z,b)&=&\nonumber\\
&&\hskip-3.0cm
-\frac{3b^{2}}{256} 
+\frac{107b^{4}}{15360} E_{0} 
-\frac{29b^{6}}{92160} E_{0}^{2} 
+\frac{b^{8}}{442368} E_{0}^{3} 
   \ , \nonumber
\end{eqnarray}
and an  examination of the extended JT supergravity expressions~(\ref{eq:V11}) and ~(\ref{eq:V21}) shows that at each order in $E_0$, the highest order in $b$ terms match the terms generated by our simple model, starting with $W_{g,n}^{\rm Bessel}$ at lowest order, through to $W_{g,n}^{\rm Airy}$ at highest. Crucially   this also works  for $g{=}3$ (comparing to $V_{3,1}^{ {\rm JT}\,({\cal N}{=}2)}$, which was computed in ref.~\cite{Ahmed:2025lxe}), and it likely persists  to higher~$g$. {\it This is all consistent with the fact that our  model captures the essence of  the low energy (BPS) sector of the ${\cal N}{\geq}2$ cases. }

Note that at $g{=}3$ (and likely beyond), it is not all of  $V_{g,1}^{ {\rm JT}\,({\cal N}{=}1)}$ that appears at zeroth order in $E_0$---it is just the Bessel piece that makes the match, connecting to  our special model. Note also that for the small ${\cal N}{=}4$ theory, for which some of the $V_{g,n}$ were also computed in ref.~\cite{Ahmed:2025lxe}, there is no such match to the terms in (\ref{eq:little-volumes}), but all terms have powers of $J$ (the $SU(2)_{\rm R}$ spin) in them, so since it is only at $J{=}0$ that there are BPS states for that model, a contradiction is avoided. See a bit later for more discussion of this case.

We've  learned the fascinating lesson that our special model can  start out life as part of    a large $N$  random matrix model with the usual $1/N$ expansion, but then turning on a finite fraction ${\widetilde\Gamma}{=}\Gamma/N$ of BPS states allows, at low energy, a self-contained sub-sector to be described in a $1/\Gamma$ topological expansion. In this sense it is a $\Gamma{\times}\Gamma$ random matrix  model associated to the BPS sector, and the associated random matrix statistics makes manifest their strong chaotic nature. With its square root edge, the model  naively resembles the   Airy model, and indeed it has such features at highest order in $E_0$, but at each order in the $1/\Gamma$ topological expansion there is a series of corrections corresponding to non-zero intersections involving mixed $\psi$-- and $\Theta$--class terms.

Moving away from the low energy limit where our special matrix model emerged, the rest of the matrix model (describing {\it e.g.,} the wider non-BPS chaotic physics of the system) joins the physics. 
Let us see next precisely how  this works, and why the special BPS model is universal among the various known matrix model descriptions of different variants of JT supergravity.

\section{The Model: Full Definition}
\label{sec:full-def}
In this supersymmetric context, the goal is to study random matrix models (RMMs) that in the double scaling limit  yield JT sueprgravity theories, but then further focus on the lowest energy sector  in order to isolate the BPS physics.
The natural class of RMM  to study is one of positive matrices $H{=}M^\dagger M$ where $M$ is a random complex matrix of size $N {\times} (N{+}\Gamma)$ (rows by columns). The matrices~$H$ are then $(N{+}\Gamma) {\times} (N{+}\Gamma)$,  and have~$\Gamma$ zeros, by design. These generically are Wishart type~\cite{10.2307/2331939} models, the prototype of which is a Gaussian model. Models of this type with higher order potentials were studied later, and multicritical potentials analogous to the multicritical ordinary Hermitian matrix models defined in refs.~\cite{Morris:1990bw,Morris:1991cq,Dalley:1991qg}.\footnote{See also refs.~\cite{Anderson:1991ku,Myers:1991akt,Lafrance:1993wy} for related work using the direct rectangular matrix approach.}
These have been used in recent times to build RMM descriptions of ${\cal N}{=}1$ JT supergravity~\cite{Stanford:2019vob,Johnson:2020heh,Johnson:2020exp}, but our focus here will be on the extended supergravity applications. The core point is that Gaussian Wishart-type model is a prototype for the universal model that appears in the low energy limit, capturing the BPS states, the gap, and a family of special non-BPS states just above the gap.

A framework within which to succinctly describe the double-scaled spectral physics uses the  quantities that arise from formulating the random matrix model in terms of orthogonal polynomial quantities~\cite{Bessis:1980ss}. An auxiliary quantum mechanical system on ${\mathbb R}$ emerges, with Hamiltonian~\cite{Gross:1990aw,Banks:1990df}:
\begin{equation}
\label{eq:aux-hamiltonian}
    {\cal H}=-\hbar^2\frac{\partial^2}{\partial x^2}+u(x)\ ,
\end{equation} and potential $u(x)$ is the double scaling limit of (a combination of) the recursion coefficients. The function $u(x)$ satisfies the defining equation of the matrix model, the ``string equation'', a  particular ODE. In this case it is:\footnote{See {\it e.g.,} refs.~\cite{Morris:1990bw,Morris:1991cq,Dalley:1991qg,Dalley:1992br}. The derivation of this equation was recently reviewed, in the current context, in ref.~\cite{Johnson:2024tgg}.}
\begin{equation}
    \label{eq:big-string}
    u{\cal R}^2-\frac{\hbar^2}{2}{\cal R}{\cal R}^{\prime\prime}
    +\frac{\hbar^2}{4}({\cal R}^\prime)^2=\hbar^2\Gamma^2\equiv{\widetilde\Gamma}^2\ ,
\end{equation}
with ${\cal R}{\equiv}\sum_{k=1}^\infty t_k R_k[u]+x$, and the objects $R_k[u]$ are polynomials in $u$ and its derivatives in the form $R_k[u]{=}u^k{+}\cdots{+}\#u^{(2k-2)}$, with the ellipsis denoting mixed derivative and non-derivative terms ($u^{(n)}$ means the $n$th $x$-derivative of $u$).\footnote{These $R_k[u]$ 
are the Gel'fand-Dikii polynomials~\cite{Gelfand:1975rn,Gelfand:1976B}, in a particular normalization.} 

Specifying the set $\{t_k\}$ of coefficients  defines the different supergravity models desired. For ${\cal N}{=}1$ models, the presence of $\Gamma$ is suppressed in the  leading order solution for $u_0(x)$, and the solution is $u_0(x){=}0$ for $x>0$ (as is evident by the fact that the term involving $\Gamma$ vanishes as $\hbar{=}0$. It was noticed in ref.~\cite{Johnson:2023ofr} that for capturing extended supergravity models with a large number of BPS states that scale with the large value $\e^{S_0}{=}\hbar^{-1}$ appropriate to black holes, it is natural to hold fixed ${\widetilde\Gamma}{=}\hbar\Gamma$ as $\hbar{\to}0$.  The~${\widetilde\Gamma}^2$ term on the right of equation~(\ref{eq:big-string}) then survives as a classical piece, and at leading order and at large $x$ there is the behavior:
\begin{equation}
    \label{eq:leading-universal}u_0(x)=\frac{{\widetilde\Gamma}^2}{x^2}\ .
\end{equation} This leading form of $u_0(x)$, which manifestly does not depend on the $\{ t_k\}$, defines a {\it universal} sector  of  all the known extended 
supersymmetric JT constructions.

To see more about why this model captures the low energy physics (including the BPS sector), it is prudent to first have a reminder of  how, in general, the energy spectrum emerges from the Hamiltonian~(\ref{eq:aux-hamiltonian}).
With some given $u(x)$ solving the string equation and hence defining~$\cal H$,
the exact ensemble average partition function computed by the matrix model is:
\begin{equation}
\label{eq:average-Z}
    \langle Z(\beta)\rangle=\int_0^\mu \langle x|\e^{-\beta\cal H}|x\rangle dx\ , 
\end{equation}
and from this (after inserting a complete set of energy eigenstates) the exact spectral density of eigenvalues of the model can be written as:
\begin{equation}
    \label{eq:exact-density}
    \rho(E)=\int_0^\mu\psi(E,x)^2dx\ .
\end{equation} At leading order, inserting the WKB form of the wavefunction $\psi(E,x)$ gives:
\begin{equation}
\label{eq:abel-projection}
    \rho_0(E)=\frac{1}{2\pi\hbar}\int_{-\infty}^\mu\frac{dx}{\sqrt{E-u_0(x)}}=\frac{1}{2\pi\hbar}\int_{E_0}^{E}\frac{f(u_0)du_0}{\sqrt{E-u_0}}\ ,
\end{equation}
(keeping only the real part of the square root)
where in the second form, the definition $u_0(\mu)\equiv E_0$ was  used and $-f(u_0){=}{-}dx/du_0$ is the Jacobian encountered in going from $x$ to $u_0$.

For the leading large $x$ form of $u_0(x)$ given in~(\ref{eq:leading-universal}):
\begin{equation}
\label{eq:abel-projection2}
    \rho_0(E)=\frac{\widetilde\Gamma}{4\pi\hbar}\int_{E_0}^{E}\frac{du_0}{u_0^\frac32\sqrt{E-u_0}}\ ,
\end{equation}
and performing the integral directly yields the expression given in equation~(\ref{eq:leading}), where the ratio $\frac{\widetilde\Gamma}{\sqrt{E_0}}\equiv\mu$. 
 
\begin{figure}
     \centering
     \includegraphics[width=0.98\linewidth]{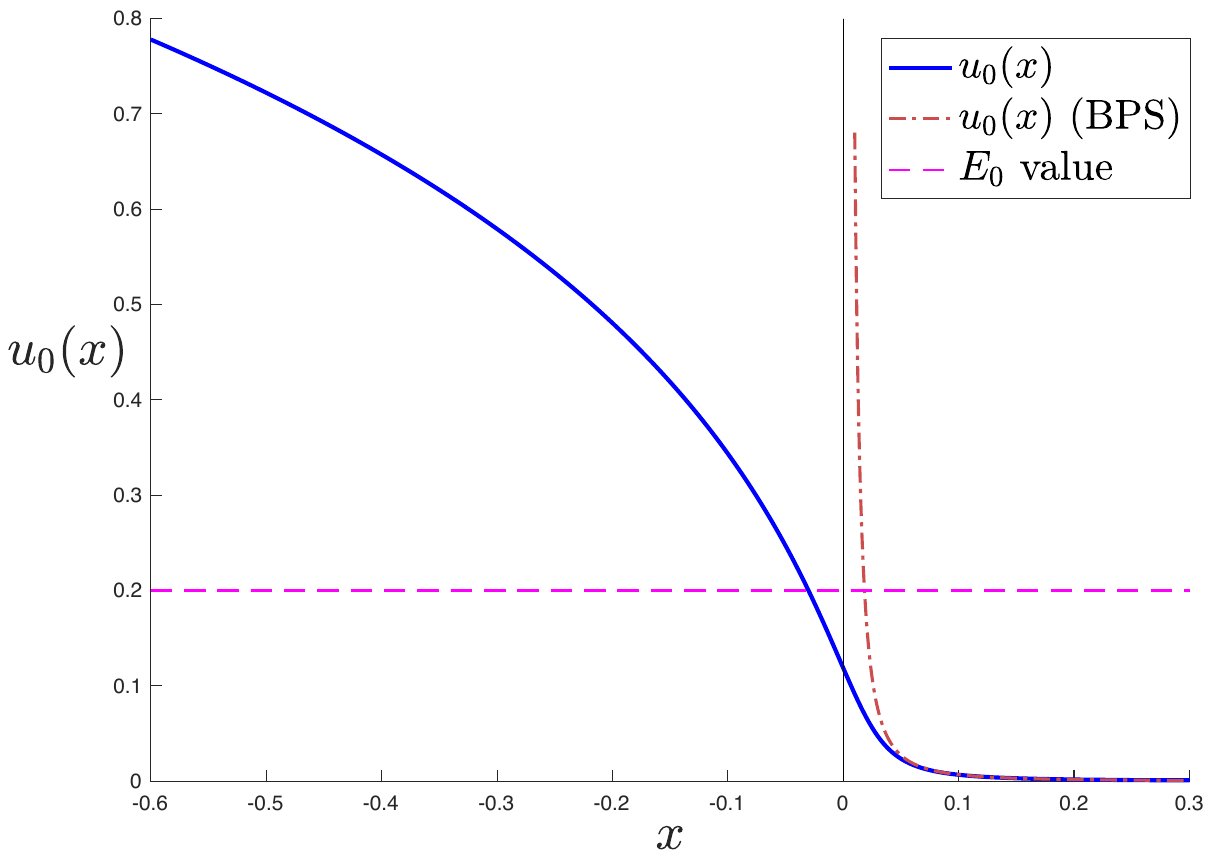}
     \caption{An example of $u_0(x)$. This is for ${\cal N}{=}2$ JT supergravity, with $E_0{=}0.2$. The BPS model curve $u_0{=}{\widetilde\Gamma}^2/x^2$ is shown as a red dashed line asymptote.}
     \label{fig:u-example}
 \end{figure}

 The following generic features are present for any of the RMMs that capture the extended JT supergravity living at the throats of BPS black holes. The full leading potential, $u_0(x)$, as a function of $x$, grows to the left (negative~$x$) passes through $x{=}\mu$ at value $u_0(\mu){=}E_0$, and falls to the far right (positive~$x$) as given in equation~(\ref{eq:leading-universal}). See the solid (blue) curve in figure~\ref{fig:u-example}, as an ${\cal N}{=}2$ JT supergravity example.\footnote{Such curves are explored in refs.~\cite{Johnson:2023ofr,Johnson:2025oty}.}
 The leading spectrum of eigenstates of the Hamiltonian $\cal H$ runs from $0$ to infinity, but  the actual RMM spectrum is constructed {\it via}~(\ref{eq:abel-projection2}), giving the continuum piece for $E\ge  E_0$. (The residue of the pole at $E{=}0$ yields the $\Gamma$ BPS states~\cite{Johnson:2025oty}.)
 
 In order to study lower and lower  energies, one goes to larger and larger $x$, and so the minimal description of the BPS sector arises from  the leading large $x$ behaviour of the potential $u_0(x)$ which is {\it universally} of the form given in equation~(\ref{eq:leading-universal})! See the dashed (red) curve in figure~\ref{fig:u-example}, showing it as an asymptote.

A quick way to find out, to all orders in $\hbar$, what is universal among the different supergravity models ({\it i.e.,} $\{t_k\}$ independent) is to simply set all $t_k{=}0$ in the full string equation~(\ref{eq:big-string}) and see what is left. 
The resulting equation  and its {\it unique} solution for $u(x)$ is:
\begin{equation}
    \label{eq:exact-special-model}
    ux^2+\frac{\hbar^2}{4}={\widetilde\Gamma}^2\ ,\quad u(x)=\frac{{\widetilde\Gamma}^2}{x^2}-\frac{\hbar^2}{4x^2}\ ,
\end{equation}
in other words, for this solution all $u_{2g}{=}0$ for $g{>}1$ and there are no non-perturbative corrections to $u(x)$. (In fact these two orders in $u(x)$  featured in Appendix~\ref{app:correction}'s computation of the $W_{g,1}(z)$ of~(\ref{eq:Wgn-interpolator}).) This is the fully-corrected low-energy (large~$x$) model common to all extended supergravity models.

It is worth noting here that for this potential, the eigenvalue equation of~${\cal H}$ turns out~\cite{Carlisle:2005wa} to simply be Bessel's equation in the variable $y{=}x\sqrt{E}/\hbar$, with solutions $J_\Gamma(y)$, giving wavefunctions $\psi(E,x){=}2^{-\frac12}\hbar^{-1}x^\frac12J_\Gamma(x\sqrt{E}/\hbar)$. The resulting {\it exact} spectral density
is, using~(\ref{eq:exact-density}):
\begin{equation}
    \label{eq:exact-density-special}
    \rho(E)= \frac{\mu^2}{4\hbar^2}\left(J_\Gamma^2(\xi)+J_{\Gamma+1}^2(\xi)-\frac{2\Gamma}{\xi}J_\Gamma(\xi) J_{\Gamma+1}(\xi)\right)\ ,
\end{equation}
where $\xi{\equiv}\frac{\mu\sqrt{E}}{\hbar}$, (a form originally derived in ref.~\cite{doi:10.1063/1.530157}). Notice again that $\mu$ and $\hbar$ appear only in the combination ${\tilde\hbar}{=}\hbar/\mu$. This  is important, since it shows (see discussion around~(\ref{eq:new-parameter})) that the observation that our special model is a $\Gamma{\times}\Gamma$ matrix model is not just an artifact of perturbation theory, but is also true non-perturbatively!

Turning on the rest of the supergravity theory through non-zero $\{t_k\}$ adds to $u(x)$  an infinite series of terms that are of order $1/x^3$ or higher inverse powers, at leading order and also  higher powers of $\hbar$. (See  the next section for a simple example.)

Given  a choice of non-zero  $\{t_k\}$, it is instructive to write the leading spectral density for general $u_0(x)$ (not just the low energy (large $x$) sector) as a $u_0$ integral, computing the Jacobian for the general ($\hbar{=}0$) string equation~(\ref{eq:big-string}). This was done in ref.~\cite{Johnson:2025oty} with the result:
\begin{eqnarray}
   && \rho_0(E) = \frac{1}{2\pi\hbar}\int_{E_0}^E\frac{\left(\sum_k kt_ku_0^{k-1}\right)du_0}{\sqrt{E-u_0}}\\
    &&\hskip4.5cm+\frac{{\widetilde\Gamma}}{4\pi\hbar}\int_{E_0}^E
    \frac{du_0}{u_0^\frac23\sqrt{E-u_0}}\ .\nonumber
\end{eqnarray}
Note the interesting fact that regardless of the functional form of $u_0(x)$, the last term is of the same form of our universal model of the low energy sector!! In the light of what has emerged in this paper so far, this is to be interpreted as the fact that the  full non-BPS spectrum of the model is made up of a mixture of parts that involve contributions from the $t_k$ sectors and the special non-BPS tail of the universal matrix model.\footnote{In retrospect, this is already clear from the original construction of ref.~\cite{Johnson:2023ofr} that required miraculous cancellations between the two sectors.}

Crucially, the full model and the universal model differ in their values of $\mu$. In general $\mu$ is, by definition, the value of $x$ at which $u_0$ takes the value $E_0$. For a general model, $\mu$'s value depends on the $\{t_k\}$ such that it precisely plays the role of what might be called  $t_0$. When we restrict to the special universal model, all the $t_k$ are zero, and then $\mu{=}\frac{{\widetilde\Gamma}}{\sqrt{E_0}}$, as seen before. It is different, in essential ways, from the value it takes when the $t_k$ are non-zero. In fact, $\mu$'s behaviour for the special model is very instructive. 

It is worth looking at two key examples, for which the full supergravity $\mu$'s were explicitly worked out in ref.~\cite{Ahmed:2025lxe}. For ${\cal N}{=}2$ JT supergravity, the value of $\mu$ is:
\begin{equation}
    \mu = \frac{J_0(2\pi\sqrt{E_0})}{2\pi} \ ,\quad \text{(${\cal N}{=}2$; full model)}
\end{equation}while for our  special model of the BPS sector\cite{Johnson:2023ofr}:
\begin{equation}
    \mu=\frac{\sin(2\pi\sqrt{E_0})}{4\pi^2\sqrt{E_0}} \ ,\quad \text{(${\cal N}{=}2$; BPS model)}
\end{equation} 
(For illustration, looking back at the ${\cal N}{=}2$ example in figure~\ref{fig:u-example}, one can read off the two different $\mu$'s by seeing where the $u_0(x)$ curves reach the value $E_0{=}\frac15$.)
The BPS sector disappears when $E_0$ grows to $\frac14$, and correspondingly the height of the peak in the model goes to zero in that case.  The spectrum simply starts at $E_0$ with purely non-BPS states. Meanwhile for small\footnote{The small and large ${\cal N}{=}4$ cases are distinguished by how the R-symmetry group is realized out of  the available global $SO(4){\sim} SU(2){\times} SU(2)$. The small case embeds it into a single $SU(2)$. See further explanation in ref.~\cite{Heydeman:2025vcc}.} ${\cal N}{=}4$ JT supergravity, the value of $\mu$ is:
\begin{equation}
    \mu=\frac{8\pi}{3}J_1(2\pi J)\ ,  \quad \text{(small ${\cal N}{=}4$; full model)}
\end{equation} while, naively,  the BPS model value is $\mu{=}\frac{1}{J}$ (following from ref.~\cite{Johnson:2024tgg}). However, in this model the BPS sector  only occurs at $J{=}0$, whereupon we see that the special model's distribution retreats to the origin (since $E_0{=}J^2{\to}0$, and grows to  infinite height (and zero width).  Large ${\cal N}{=}4$, on the other hand,  was shown/argued in ref.~\cite{Johnson:2025oty} to be described (as a matrix model) in terms of a pair of ${\cal N}{=}2$ models, and so its features are more akin to what we've already seen for~${\cal N}{=}2$.

\section{The simplest embedding example}
As stated above, the special model is a natural description of the low energy sector of a wide range of matrix models that differ by the  choice of $\{t_k\}$). It is illustrative to explore  the simplest non-trivial embedding of it into a larger model,  coming from turning on just  $t_1$. This takes the basic Wishart model and adds gravity by including  $k{=}1$ multicritical behaviour to it~\cite{Morris:1992zr,Dalley:1991qg}. 

In this case, writing ${\cal R}{=}t_1 u_0{+}x$  in (\ref{eq:big-string}) gives a second order ODE.\footnote{In this case, the ODE is actually equivalent to the Painlev\'{e}~XXXIV equation~\cite{InceBook}.}
Putting $\hbar{=}0$, the leading   equation is simply a cubic in $u_0(x)$:
\begin{equation}
\label{eq:cubic}
    u_0(t_1u_0+x)^2={\widetilde\Gamma}^2\ ,
\end{equation}
with real solution:
\begin{eqnarray}
    u_0&=& \frac{1}{3 \mathit{t_1}}\left(\frac{{D}^{\frac{1}{3}}}{2}+\frac{2 x^{2}}{{D}^{\frac{1}{3}}}-2 x\right)\ , \quad\text{where:}\\
  D&=& \left(12 \sqrt{3}\, {\widetilde\Gamma} \mathit{t_1}\sqrt{27 {\widetilde\Gamma}^{2}  +\frac{4 x^{3}}{\mathit{t_1}}}\,  +108 {\widetilde\Gamma}^{2} \mathit{t_1} +8 x^{3}\right)\ , \nonumber
\end{eqnarray}
upon which expanding around large positive $x$ yields:
\begin{equation}
  u_0= \frac{{\widetilde\Gamma}^{2}}{x^{2}}-\frac{2 {\widetilde\Gamma}^{4} t_1}{x^{5}}+\frac{7 {\widetilde\Gamma}^{6} t_1^{2}}{x^{8}}-\frac{30 {\widetilde\Gamma}^{8} t_1^{3}}{x^{11}}\cdots\ ,   
\end{equation}
showing the universal piece that defines the special model as the first term, followed by corrections that fall off even faster at large $x$. Similarly, $\hbar^2$ corrections to $u(x)$ start as:
\begin{equation}
    u_2(x)=-\frac{1}{4 x^{2}}+\frac{5 {\widetilde\Gamma}^{2} t_{1}}{x^{5}}-\frac{217 {\widetilde\Gamma}^{4} t_{1}^{2}}{4 x^{8}}+\frac{500 {\widetilde\Gamma}^{6} t_{1}^{3}}{x^{11}}+\cdots
\end{equation}
and there are non-zero contributions generated by $t_1$ for all the higher $u_{2g}(x)$.
The leading spectral density can be readily solved for as:
\begin{eqnarray}
    &&\rho_0(E) = \frac{1}{\pi\hbar}t_1{\sqrt{E-E_0}}+\frac{1}{2\pi\hbar}\frac{\widetilde\Gamma}{\sqrt{E_0}}
    \frac{\sqrt{E-E_0}}{E}\ ,
\end{eqnarray}
where $E_0$ is determined by the choice of $\mu$, which isn't  fixed by anything here. The point here is  that, the $E{>}E_0$ energy spectrum has a natural division into a component coming from the rest of gravity (first part) and a universal special component that is, as we've seen, correlated with the BPS sector. Their relative admixtures depend on the relative strength of $t_1$ and ${\widetilde\Gamma}$. 

\section{LMRS and the Jacobi Ensemble}
\label{sec:LMRS-Jacobi}
While the energy spectrum of the $\Gamma{\times}\Gamma$ matrix model,  and all its correlation functions, have been argued to be telling us (perhaps  indirectly) about the $\Gamma$ BPS states, this isn't the only thing we can capture about the model. Indeed, as emphasized by Lin, Maldacena, Rozenberg and Shan (LMRS)~\cite{Lin:2022rzw,Lin:2022zxd}, since the BPS states all have zero energy, the statistics of the random Hamiltonian's energies will not be of any use in directly seeing chaos for that sector. One can instead pick a fixed operator ${\cal O}$ in the theory of interest and project it onto the BPS sector. The matrix elements $R_{ij}$ of that  projected operator are then  argued, essentially through the eigenvalue thermalization hypothesis (ETH)~\cite{Deutsch:1991msp,Srednicki:1994mfb,Rigol:2007juv,DAlessio:2015qtq}, to have random matrix statistics.

Such a reduction can be done in our universal model, and it is natural to do it in the full (non-double-scaled) parent Wishart-type model whose construction was discussed at the beginning of Section~\ref{sec:full-def}. Since the rectangular $M$ constructed above naturally annihilates $\Gamma$ orthogonal vectors ({\it i.e.,} $(N+\Gamma)\times1$ column vectors), 
we can normalize them to unit length and form an orthonormal basis with them, arranging them in columns to make $U$, an $(N{+}\Gamma)\times\Gamma$ matrix. The $(N{+}\Gamma)\times(N{+}\Gamma)$ matrix $P_{\rm BPS}=UU^\dagger$ is then a projector into the BPS sector. So for a  fixed $(N{+}\Gamma)\times(N{+}\Gamma)$ matrix~$O$, $P_{\rm BPS}OP_{\rm BPS}$  is its restriction to the to the BPS subspace. So the $\Gamma{\times}\Gamma$ matrix that's  the LMRS diagnostic  is:
\begin{equation}
    R_O=U^\dagger O U\ ,
\end{equation}
and we should study the statistics of its eigenvalues  across the random $M$ ensemble. What matrix to use for $O$ is a choice, and a natural one suggests itself. Pick $O$ to be a projection matrix of rank $\Gamma$, {\it e.g.,} $O{=}{\rm diag}(\bm{1}_\Gamma,\bm{0}_N)$. It projects to a fixed $\Gamma$-dimensional plane within the full space on which $H$ acts. Since the BPS subspace is itself essentially a (random) $\Gamma$-dimensional plane, the eigenvalues of the matrix $R_O$ are simply the set $\{\cos^2\theta_i\}$, where $i{=}1,\ldots,\Gamma$, and the $\theta_i$ are the principal angles giving the relative orientation of $O$'s plane to the random BPS plane. The eigenvalues therefore live between zero and one, and such systems are in fact well-studied. For Gaussian distributed $M$, the joint distribution of eigenvalues is the Jacobi distribution~\cite{ForresterBook,Collins:2005}, and indeed it has classic random matrix (chaotic) features, such as eigenvalue repulsion, {\it etc}. In the large $N$ limit with $\widetilde\Gamma{=}\Gamma/N$ fixed (and $\Gamma<N$), its normalized leading spectral density is:
\begin{equation}
 \rho^J_{0}(\lambda)
 =
 \frac{1+\widetilde\Gamma}{2\pi\widetilde\Gamma}\,
 \frac{\sqrt{\lambda_+-\lambda}}
      {\sqrt{\lambda}(1-\lambda)}\ ,
 \qquad
 \lambda_+
 =
 \frac{4\widetilde\Gamma}
      {(1+\widetilde\Gamma)^2}\ .
      \label{eq:jacobi-density}
\end{equation}
It has a hard edge divergence at zero (representing the fact that it is common to find orthogonality between dimensions of the two planes)  and ends at a soft edge $\lambda_+{<}1$. The distribution above is readily confirmed in our numerical experiments discussed in the next section. 

This is a very interesting companion model of the BPS physics, complementing our study of the spectral features of $H$ itself. More features are worth studying (such as changing the rank of the fixed projector $O$), but this will be left for a future project.

\section{Fun with Random Matrices}
The essential features  of the universal model arise as a limit of the classic Wishart random matrix model and so one can simply build it explicitly as a laboratory for exploration.
Here's how to do it, using as an example $\Gamma{=}10$ and  $N{=}100$, working  in {\tt MATLAB}. (The lines of a sample program are given in Appendix~\ref{app:matlab-program}.) It is  straightforward to generate $110{\times} 110$ Gaussian random matrices~$M$, after which  ten  rows can be deleted, making $M$ a $100{\times}110$ rectangular matrix. Forming $H{=}M^\dagger M$ makes a $110{\times}110$ square matrix with $\Gamma{=}10$ zero eigenvalues. One can then compute the eigenvalues of such a randomly generated matrix and store these data. Then repeat this some number of times, say $10^5$, for good statistics. 

\begin{figure}[t]
    \centering
    \includegraphics[width=0.98\linewidth]{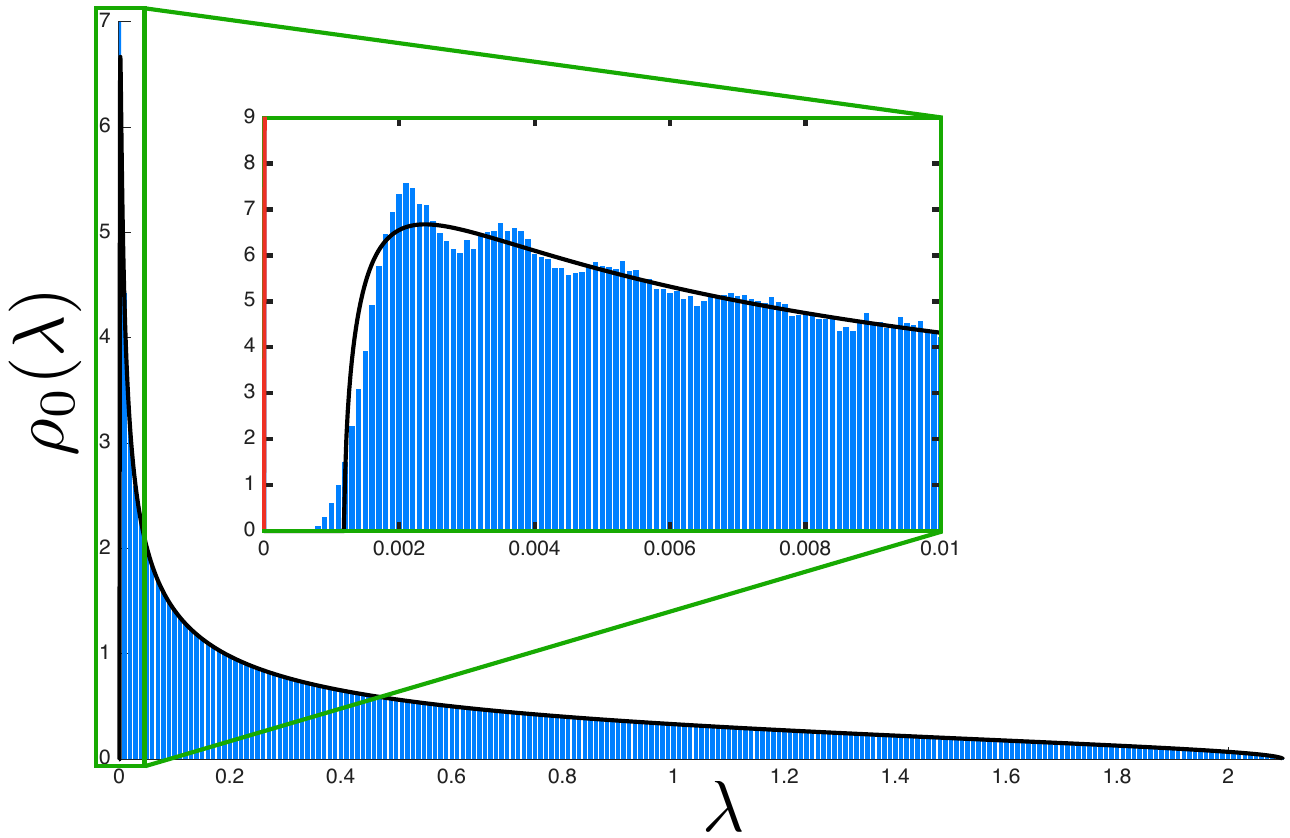}
    \caption{Histogrammed statistical data, across $10^5$ samples, from a Wishart-type model with $\Gamma{=}10$, $N{=}100$, scaled into the endpoint. The Marchenko-Pastur leading distribution~(\ref{eq:marchenko-pastur}) emerges (smooth curve). The inset shows the area close to the origin where the gap in the leading distribution is visible, with the continuum starting at $\lambda_-{\simeq}0.001191$.}
    \label{fig:BPS-numerics-1}
\end{figure}

One can then histogram the data. To match the results to the classic Marchenko-Pastur\cite{Pastur:1967zca} distribution that appears at leading large $N$, some scaling is needed. The distribution is (writing $\Gamma/N={\widetilde\Gamma}$):
\begin{equation}
\label{eq:marchenko-pastur}
    \rho^{\rm MP}_0(\lambda) =  
   \frac{\sqrt{(\lambda_+-\lambda)(\lambda-\lambda_-)}}{\pi\lambda}\ ; \quad \lambda_\pm=\frac{\widetilde\Gamma}{2}+1\pm\sqrt{{\widetilde\Gamma}+1}\ ,
\end{equation}
and the $\lambda$ here is for potential $V(\lambda){=}{-}2\lambda$ and there is  an~$N$ multiplying the overall potential in the matrix model action for matrices. On the other hand, a numerical Gaussian generator for $M$ has standard normalization $\e^{-{\tilde y}^2/2}$, where ${\tilde y}$ are~$M$ eigenvalues. So this translates into $2N\lambda{=}{\tilde y}^2/2$ and hence the $H$ eigenvalues ${\tilde\lambda}{=}{\tilde y}^2$ in the  data should be scaled to $\lambda{=}{\tilde\lambda}/(4N)$, and then histogrammed. The successful results are shown in figure~\ref{fig:BPS-numerics-1}.

The universal physics of interest is near the left endpoint, working in a limit where we keep both $\lambda$ {\it and} the $\lambda_-$ small while retaining a finite gap. We send ${\widetilde\Gamma}{\equiv}\frac{\Gamma}{N}{\to}\epsilon{\widetilde\Gamma}$, which amounts to now defining $\hbar$ through $\hbar{=}1/(N\epsilon)$, which  is kept finite by sending  $\epsilon\to0$ while $N\to\infty$. So according to~(\ref{eq:marchenko-pastur}),  $\lambda_-{=}\frac{\epsilon^2}8{\widetilde\Gamma}^2+\cdots$.  
Meanwhile under the square root, we have $(\lambda_+{-}\lambda)(\lambda-\lambda_-)\simeq(\lambda_+{+}\lambda_-)\lambda{-}\lambda_+\lambda_-{=}(\epsilon{\widetilde\Gamma}+2)\lambda{-}\epsilon^2{\widetilde\Gamma}^2/4$, keeping only linear $\lambda$  terms in anticipation that it is small. So defining $\epsilon^2E{=}4(\epsilon{\widetilde\Gamma}+2)\lambda$, the energy scale $E$ governs the physics in the scaling limit. In this variable the gap is at $E_0={\widetilde\Gamma}^2$, and: 
\begin{equation}
    \rho_0(E) = \frac{1}{2\pi\hbar}\frac{\sqrt{E-{\tilde\Gamma}^2}}{ E}\ ,
\end{equation}
{\it i.e.,} the special model of this paper with $\mu{\equiv}{\widetilde\Gamma/\sqrt{E_0}}{=}1$. To get the $\mu$ dependence, choose instead, by hand, a scale~$\mu$ such that  $\epsilon^2E{=}4({\widetilde\Gamma}+2)\lambda/\mu^2$ gives gap $E_0{=}{\widetilde\Gamma}^2/\mu^2$, and a factor $\mu$ in front, recovering  the full model.

\begin{figure}[t]
    \centering
    \includegraphics[width=0.98\linewidth]{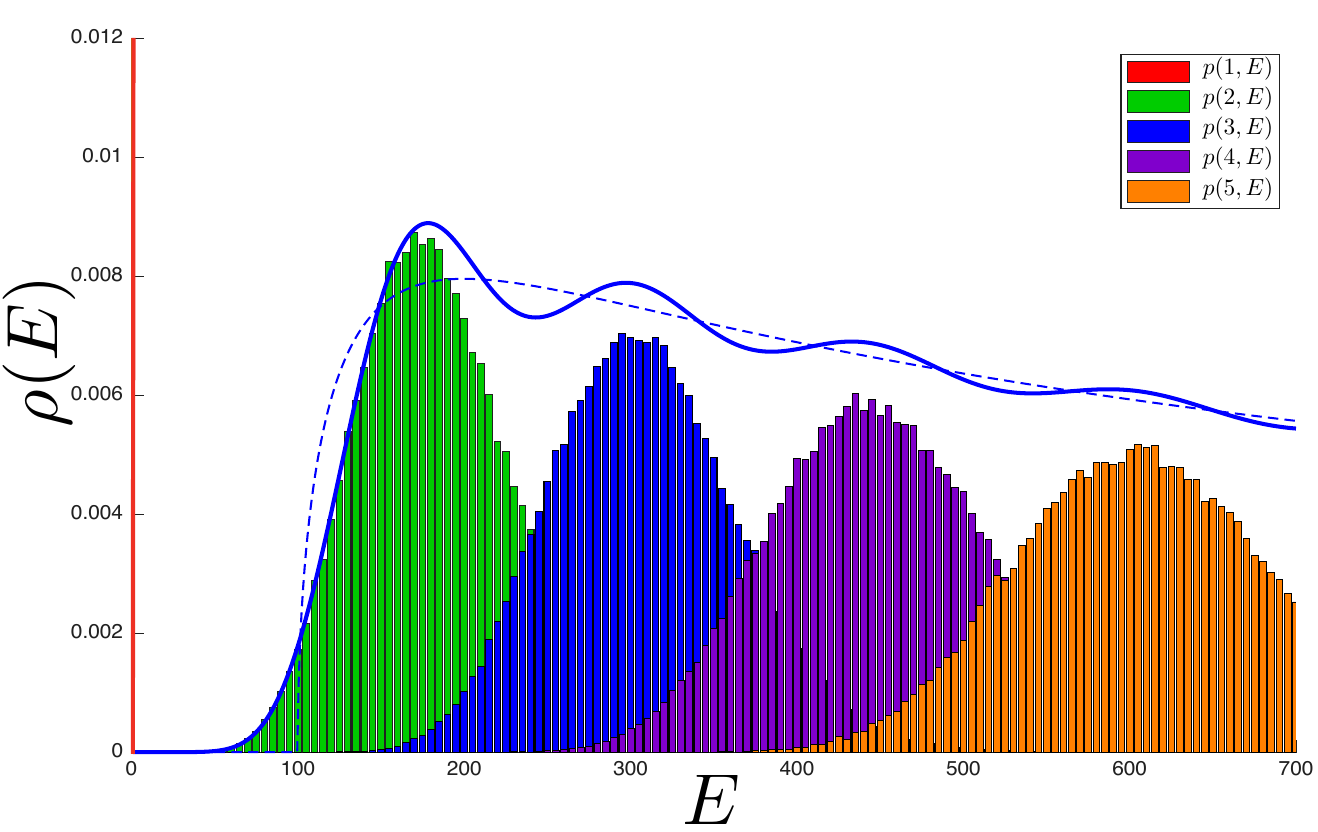}
    \caption{The histogrammed data from figure~\ref{fig:BPS-numerics-1}, zoomed into the edge region, by scaling the energies as described in the text. Now the continuum begins at $E_0{=}{\widetilde\Gamma}^2=100$, with~$\hbar$  effectively set to unity. Shown in red are the BPS states, and distributions 
    of the first four non-zero energy levels, across~$10^5$ samples. Curves showing the analytic  expression for both the leading ($\rho_0(E)$, dotted) and exact ($\rho(E)$, solid)  model are superimposed on top.  
    }
    \label{fig:BPS-numerics-2}
\end{figure}

For handling the data in the numerical work, this suggests one final scaling choice that will allow a direct comparison to the known exact results in terms of Bessel functions $J_\Gamma(E)$, for integer $\Gamma$. The ${\widetilde\Gamma}{=}10$ data can be turned into $\Gamma{=}10$ data, effectively setting $\hbar{=}1$, by scaling $\lambda$ in the data by a factor of $N^2$, to get data that is now $E{=} 4({\widetilde\Gamma}+2)N^2\lambda$. The results are shown in figure~\ref{fig:BPS-numerics-2}, along with the analytic expressions for $\rho_0(E)$ and the exact $\rho(E)$. The  results from the numerics for the ground states and first four energy levels are plotted on top as histograms. The bumps in $\rho(E)$ originate from summing the individual probability peaks for energy levels, as is illustrated nicely here. (Such peaks can be computed using non-perturbative techniques in the matrix model, as illustrated in the gravity context in ref.~\cite{Johnson:2021zuo}.)

It is worth emphasizing that the basic Wishart model here naturally leads to  the value $\mu{\equiv} {\widetilde\Gamma}/{\sqrt{E_0}} {=}1$, {\it i.e.,}  $E_0{=}{\widetilde{\Gamma}}^2$. A different value for $\mu$ {\it can} be introduced by hand (as noted), but there's no physical significance for it in Wishart on its own.   However, as is now hopefully clear, the parameter $\mu$ is crucial for the model's usefulness in capturing the BPS sector physics embedded in a larger system.

\begin{widetext}
  From here, it is straightforward to mine these data sets further, as was for example done long ago in ref.~\cite{Johnson:2021rsh} as an illustrative toy model for studying quenched {\it vs.} annealed free energy in 2D gravity.\footnote{Another  quantity of interest (particular in the chaos context) that can readily  be computed with these data  is the spectral form factor. See upcoming work~\cite{Saraswat:2026cuo} for  studies of the spectral form factor for random systems with a gap, such as this one.} The matrix $R_O$ of our LMRS-like construction in Section~\ref{sec:LMRS-Jacobi} can be readily constructed from 
  the data gathered here too, by simply constructing $U$ from each of the samples of $M$, and following the recipe. The distribution~\eqref{eq:jacobi-density} readily emerges, as shown in figure~\label{fig:BPS-numerics-3}. There, the histogrammed data are plotted alongside the exact Jacobi distribution density~\cite{Collins:2005}, which is:
  \begin{equation}
 \rho^J(\lambda)
 =
 \frac{(1-\lambda)^{N-\Gamma}}{\Gamma}
 \sum_{k=0}^{\Gamma-1}
 \left(2k+N-\Gamma+1\right)
 \left[
 P_k^{(N-\Gamma,0)}(2\lambda-1)
 \right]^2\  ,
 \qquad
 0<\lambda<1\ ,
 \label{eq:jacobi-density-exact}
\end{equation}
the leading part of which is the density~\eqref{eq:jacobi-density}. Here, the $P_k$ are the Jacobi polynomials of degree $k$. For the choice of parameters here, the classical soft edge is at $\lambda_+\simeq 0.33058$. The mean of the distribution is $\Gamma/(N{+}\Gamma){\simeq}0.09091$, which is confirmed on the plot. Further details of this and other explorations  are left for a future publication.
\begin{figure}[t]
    \centering
    \includegraphics[width=0.8\linewidth]{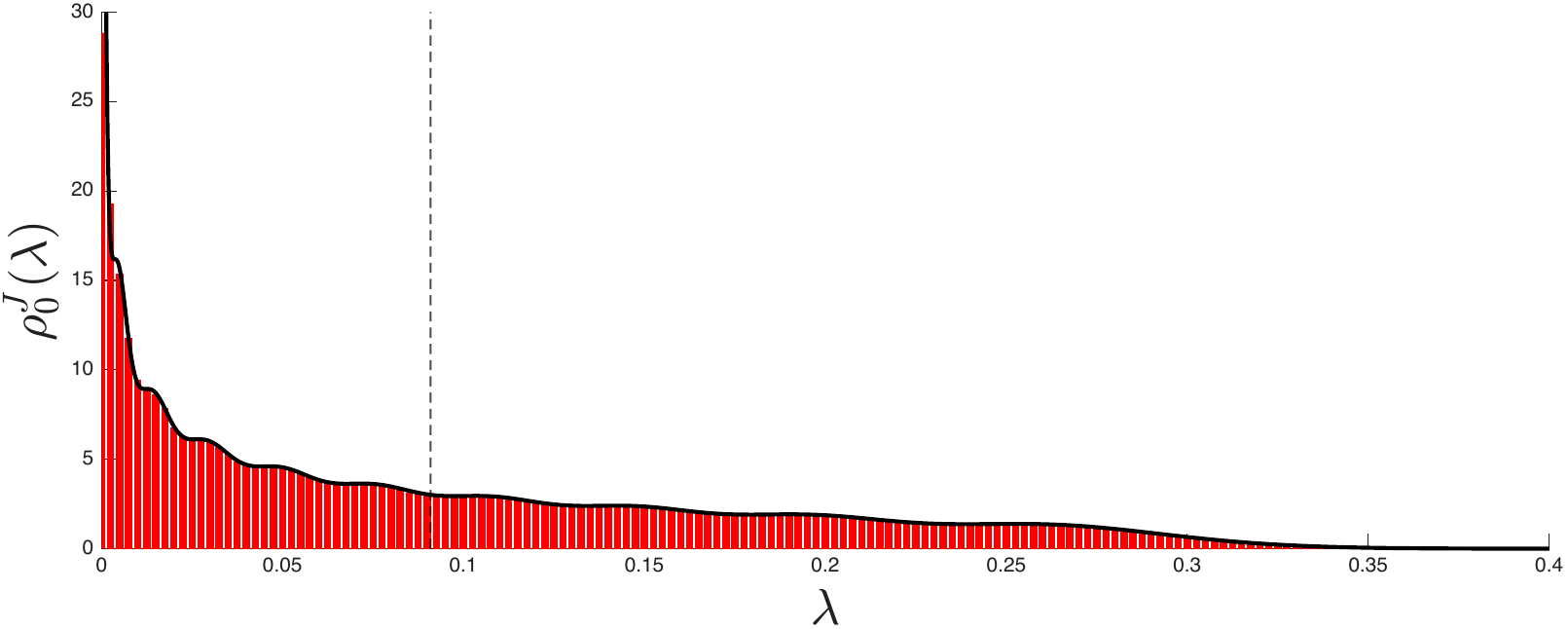}
    \caption{Histogrammed statistical data, across $10^5$ samples, from the  Jacobi model of the LMRS-like probe of Section~\ref{sec:LMRS-Jacobi}, with $\Gamma{=}10$, $N{=}100$. The exact  Jacobi density~(\ref{eq:jacobi-density-exact})  (smooth curve) is emerging. The exact mean,
$\langle\lambda\rangle=\Gamma/(N+\Gamma)=1/11\simeq0.0909$,
is also shown (dashed line). There is a hard edge at the origin and, in the leading large-$N$ limit, ends at the soft edge $\lambda_+{=}4\widetilde\Gamma/(1+\widetilde\Gamma)^2{\simeq}0.331$.}
    \label{fig:BPS-numerics-3}
\end{figure}
    
\end{widetext}

\section{Discussion}

For a large number, $\Gamma{\sim} \e^{S_0}$, of BPS states, where $S_0$ is the black hole entropy, we have seen that a special  $\Gamma\times\Gamma$ random matrix model emerges as a universal model of the (ultra) low energy dynamics. The model is present as a subsector within a variety of the larger $N{\times} N$ double-scaled random matrix models of the JT supergravity theories with extended supersymmetry that have been shown to govern the near-horizon throats of BPS black holes ($N{\sim}\e^{S_0}$, but $\Gamma{<}N$), but the special model is itself entirely  independent of the gravity details. 

It is proposed that this model is {\it also} a universal model  that captures features of fortuitous BPS sector of the holographically dual gauge theory (at least for large $\Gamma$ where one might hope for some universality to emerge). In particular, certain key features of the model  become predictions. Perhaps the  foremost of those is the fact that accompanying the  BPS states at $E{=}0$ are a set of non-BPS states that lurk  above the gap~$E_0$, with a density peaked at $2E_0$. They  are concomitant with the BPS states (see figure~\ref{fig:BPS-density}). 
The presence of these states might seem surprising at first, but on the other hand they fit naturally with the defining phenomenon of fortuity:  For some given set of fortuitous states already in the BPS sector, there {\it should} be some non-BPS states that are on the verge of becoming BPS. We've already learned from holography that ``on the verge'' cannot be smaller than the gap $E_0$, and so that should set the scale at which any nearby  states lurk in readiness to fortuitously appear below the gap (the ``chaos invasion'' in the terminology of ref.~\cite{Chang:2024lxt}). The higher above $E_0$ one is, the less likely such states should be, which is why the special spectrum falls off at large $E$. The model's unambiguous prediction is a characteristic $1/\sqrt{E}$ fall off.

The use of the model of this paper as an earmark of BPS chaos has resonance with the ``supercharge chaos'' recently discussed by Chang, Chen, Sia, and Yang (CCSY)  in ref.~\cite{Chang:2024lxt}. Their discussion is motivated by  Turiaci and Witten (TW)~\cite{Turiaci:2023jfa}'s observation (already reviewed in the Introduction) that ${\cal N}{=}2$ JT supergravity can be recast as a random matrix model. CCSY conjectures that TW's matrix model is a good diagnostic tool for BPS chaos. This is undoubtedly correct, but this paper makes it clear that there's far more contained in that model than is needed to do the job: There's a much simpler {\it universal} $\Gamma{\times}\Gamma$ random matrix model that lives at the tail that is the essential ingredient. In ref.~\cite{Johnson:2023ofr}'s re-formulation of TW's matrix model as an admixture of multicritical models determined by some coefficients~$\{ t_k\}$, it is easy to find the universal model. Just set all the $t_k$ to zero, and the underlying string equation still has a non-trivial solution. (See paragraph around equation~(\ref{eq:exact-special-model}).)

It is key to note that the  independence from the details of the gravity sector is consistent with a striking fact about the model. It is a topological model, which we saw interpolates between two known topological models,  Airy  and  Bessel. Like them, it evidently has an elegant description as a generator of  intersection numbers associated to the topology of the compactified moduli space of punctured Riemann surfaces. Some of its features were explored in this paper,  although there is more to be uncovered in that regard.  Connections to a special class of deformations of certain $\Theta$-classes seem relevant here, and are likely to be fruitful to explore.

Note that the tight correspondence between the $\Gamma$ BPS states at $E{=}0$ and the special non-BPS cloud at $E>E_0$ is further solidified by the intersection theory picture. It shows how turning on $E_0$  is a quite natural deformation parameter in the topological context. Overall, these results also shows a refinement of the key role of eigenvalue repulsion in naturally generating the supergravity gap that was  discussed in ref.~\cite{Johnson:2024tgg}. While (in matrix model language) the $\Gamma$ degenerate eigenvalues push away all the non-zero eigenvalues, evidently the first (random)~$\Gamma$  of them form a privileged  subsector that is peaked just above the gap.

It was also observed here that the manner in which the intersection theory data are organized (through a kind of deformation by $E_0{\sim}\hbar^2\Gamma^2$) is an avatar of an organization that was observed for the curious structure of Weil-Petersson volumes of ${\cal N}{=}2$ JT supergravity models  ~\cite{Turiaci:2023jfa,Ahmed:2025lxe}, (with similar features extending to large ${\cal N}{=}4$ as well). In fact, it is a precise model of what is going on in those larger models, which was hitherto mysterious in some regards. The point is that at least if one focuses on the volumes, such supergravity models  can be thought of as perturbations (with parameters $\{t_k\}$) of the basic spectral curve of our simple model, just as  the standard Weil-Petersson spectral  curve $({\bar x}{=}z^2,y{=}\sin z)$ can be thought of as pertubation of Airy $({\bar x}{=}z^2,y{=}z)$ in the context of ordinary JT gravity.

A cynic might say that the model that featured in this paper has been around in the literature for some time, and so dismiss the whole matter as old news. In fact, sometimes it is also called the Bessel model, but here the name was reserved for the model(s) with leading density $\rho_0(E){=} 1/(2\pi\sqrt{E})$, closer to the terminology used in the topological recursion and intersection theory literature. There's an infinite family of them indexed by~$\Gamma$, and they can be built from wavefunctions constructed from Bessel functions of order~$\Gamma$. (In fact, the $\Gamma{=}0$ and $\Gamma{=}\pm\frac12$ variants featured heavily as models of the tail of various ${\cal N}{=}1$ JT supergravity models in refs.~\cite{Stanford:2019vob,Johnson:2020heh,Johnson:2021owr}.) The model here is an ``extreme Bessel'' limit where $\Gamma$ is taken large. Nevertheless even in that limit it is still in many respects  an ``old model'', but it is being cast in a quite different role in this paper, because of a number of its features  have not been put together in this context before.

A case in point is  the parameter called $\mu$ in this paper. Normally in the literature, that is set to some standard conventional value, typically  $\mu{=}1$. This almost entirely obscures the main utility of the model in its role in this paper. Key was to keep $\mu{=}{\tilde\Gamma}/\sqrt{E_0}$ as a parameter of the model, determined by the larger supergravity theory within which it is embedded. Then  the fact that,  once   the model emerges in the low energy limit $\mu$ appears everywhere in the combination $\mu/\hbar$, makes it immediately clear that the matrix model topological coupling is effectively~${\sim}1/\Gamma$. Hence, a $\Gamma{\times}\Gamma$  matrix model has decoupled from the full supergravity RMT. This is a peculiar feature of the leading  potential $u(x){=}\hbar^2\Gamma^2/x^2$ of the model (in the Hamiltonian~(\ref{eq:aux-hamiltonian}) of the auxiliary  quantum mechanics). The kinetic term in the quantum mechanics also has an~$\hbar^2$ and so that overall factor can be scaled out entirely, leaving $1/\Gamma$ as the effective $\hbar$.  This is quite different from, say, the Airy model. While it appears at the low energy tail of ordinary JT, it is still an $N{\times}N$ matrix model since there is no $\hbar^2$ (times a counting parameter) in the potential $u(x){=}{-}x$. Finally, not setting~$\mu$ to some fixed value makes the properties of the spectrum in the leading density much more clear with regards  the underlying $\Gamma$ BPS states, as shown in figure~(\ref{fig:BPS-density}).

There are probably many other things to be learned from this model that would be useful to characterize the physics of  the (chaotic) BPS sector. Most straightforward to tackle are non-perturbative effects, easily accessible with the techniques of this paper since the model  is formulated fully non-perturbatively---It was simply a choice  to focus on the perturbative features. The string equation solution itself~(\ref{eq:exact-special-model}) has no non-perturbative corrections and so all the non-perturbative physics come from the exact wavefunctions (see text around (\ref{eq:exact-density-special}), or equivalently the Gel'fand-Dikii equation approach discussed in Appendix~\ref{app:correction}. Already clear is that the effective potential for one eigenvalue (easily computed by continuing the integral of the leading spectral density) heralds the tunneling of eigenvalues out of the leading $E{>}E_0$ region. This effect is evident from looking at the tail of the fully non-perturbative expression (\ref{eq:exact-density-special}) for the density $\rho(E)$ and the numerical results, plotted in figure~\ref{fig:BPS-numerics-2}. Eigenvalues can leak into the (classical) gap. Understanding what exactly this means for the fortuitous BPS sector should be interesting to explore.

\bigskip
\begin{acknowledgments}
CVJ thanks Maciej Kolanowski, Don Marolf, Krishan Saraswat, Mark Srednicki,  Misha Usatyuk, and Edward Witten for remarks.
CVJ  also thanks  the  US Department of Energy (\protect{DE-SC} 0011687) for  support,  and  Amelia for her support and patience.    
\end{acknowledgments}

\bigskip
\appendix

\section{Computing corrections to $\rho_0(E)$.}
\label{app:correction}
A very efficient method for computing all the perturbative corrections to the leading spectral density $\rho_0(E)$ follows from the presence of the auxiliary Hamiltonian ${\cal H}{=}{-}\hbar^2\partial_x^2+u(x)$ in the definition~(\ref{eq:average-Z}) of the (averaged) partition function over the  ensemble of Hamiltonians. As noticed in refs.~\cite{Johnson:2020heh,johnson:2024bue}, the efficiency comes from the fact that the corrections are all packaged into an ordinary differential equation (ODE), namely the one derived by  Gel’fand-Dikii~\cite{Gelfand:1975rn}  for the diagonal resolvent ${\widehat R}(E,x)\equiv\hbar\langle x|({\cal H}-E)^{-1} |x\rangle$ of ${\cal H}$:
\begin{equation}
\label{eq:gelfand-dikii}
    4(u-E)\widehat{R}^2-2\hbar^2\widehat{R}\widehat{R}^{\prime\prime}+\hbar^2\widehat(R^\prime)^2=1\ ,
\end{equation}
which is sourced by the potential $u(x)$, which separately solves a string equation (in our case, equation~(\ref{eq:big-string})).
The fact is that $\rho(E)$ can be recovered as:
\begin{equation}
\label{eq:rho-relation}
    \rho(E)=\frac{{\rm Im}}{\pi\hbar}\int_{-\infty}^\mu {\widetilde R}(E,x)dx\ .
\end{equation}
Focusing on perturbation theory, the string equation yields  a perturbative form for $u(x)$:
 \begin{equation}
 \label{eq:u-expansion}
     u(x){=}\sum_{g=0}^\infty u_{2g}(x)\hbar^{2g}+\cdots\ ,
 \end{equation}
and the same can be said for the resolvent equation:
\begin{equation}
 \label{eq:R-expansion}
     \widehat{R}(x,E)=\sum_{g=0}^\infty \hbar^{2g}\widehat{R}_g(x,E)+\cdots\ ,
 \end{equation}
(In both cases the ellipses denote non-perturbative parts.) The perturbative terms in ${\widehat R}(x,E)$ arise from the structure of the equation~(\ref{eq:gelfand-dikii}) itself as well as the form~(\ref{eq:u-expansion}), which is an input. This results in (after choosing a sign convention):
\begin{widetext}

\begin{eqnarray}
&&
{\widehat R}(x,E)
= -\frac12\frac{1}{[u_0(x)-E]^{1/2}}
+
\frac{\hbar^2}{64}\left\{
\frac{16u_2(x)}{[u_0(x)-E]^{3/2}}
+\frac{4u^{\prime\prime}_0(x)}{[u_0(x)-E]^{5/2}}-\frac{5(u^{\prime}_0(x))^2}{[u_0(x)-E]^{7/2}}\right\}
\label{eq:gelfand-dikii-A}
\\
&&\hskip+1cm
+\frac{\hbar^4}{4096}\left\{
\frac{1024u_4(x)}{[u_0(x)-E]^{3/2}}
-\frac{256[3u_2(x)^2-u^{\prime\prime}_2(x)]}{[u_0(x)-E]^{5/2}}
+\frac{64[u^{(4)}_0(x) -10u_2(x) u^{''}_0(x)-10u^\prime_0(x)u^\prime_2(x)]}{[u_0(x)-E]^{7/2}}\right. \nonumber \\
&&\hskip+0.5cm
\left.
-\frac{16[28u^{(3)}_0(x)u^\prime_0(x)+21u^{\prime\prime}_0(x)^2-70u_2(x)u_0(x)^2]}{[u_0(x)-E]^{9/2}}
+\frac{{1848}u^{\prime}_0(x)^2u^{\prime\prime}_0(x)}{[u_0(x)-E]^{11/2}}-\frac{1155u^{\prime}_0(x)^4}{[u_0(x)-E]^{13/2}}\right\} +\cdots\ .   \nonumber
\end{eqnarray} 
\end{widetext}
Comparing the leading term to~(\ref{eq:abel-projection}), one sees that $ \rho_0(E){=}\frac{{\rm Im}}{\pi\hbar}\int_{-\infty}^\mu {\widetilde R}_0(E,x)dx$  indeed matches it. The higher order terms ${\widehat R}_g(E,x)$ accordingly generate higher order corrections in $\rho(E)$.

 By simple recursively solving the string equation, all the $u_{2g}(x)$ can  be written in terms of $u_0(x)$ and its derivatives. 
 For example:
\begin{equation}
\label{eq:u2}
    u_2(x)
    =-\frac{1}{12}\frac{d^2}{dx^2}\ln(u_0')
    =\frac{u_0''^2-u_0'u_0'''}{12u_0'^2}\ .
\end{equation}
 
 The remarkable thing~\cite{johnson:2024bue} is the observation that in replacing the $u_{2g}(x)$ in~(\ref{eq:gelfand-dikii-A}) by their $u_0(x)$ (and derivatives) dependence, each order in $\hbar$ turns out to be a total derivative, {\it i.e.,} ${\widehat R}_{g}(E,x)=\frac{d}{dx}Q_g(E,x)$!  For example:
 \begin{equation}
\label{eq:totally-awesome}
    Q_1(x,E)=-\frac{u_0^{\prime\prime}(x)}{48 u_0^\prime(x)[u_0(x)-E]^{3/2}}+\frac{u_0^\prime(x)}{32[u_0(x)-E]^{5/2}}
    \ .
\end{equation}
 It turns out the $Q_g(E,x)$ vanish at $x=-\infty$ and so the corrections are then simply given by $Q_g(E,x)$ evaluated at $x=\mu$, which results only from behaviour of $u_0$ and its derivatives at $x=\mu$ (the natural ``Fermi surface'', in another language). 
Those $Q_g(E,\mu)$, after multiplying by~$-2z$ and changing variables from $u_0(\mu)-E\equiv E_0-E$ to $z^2$ are precisely the $W_{g,1}(z)$ correlation functions. This is all refined and discussed much further in ref.~\cite{Ahmed:2025lxe}, where it is applied to WP volumes for extended JT supergravity, and in ref.~\cite{Johnson:2026jbq} where there is also discussion of  intersection theory quantities, and the method extended to derive non-perturbative results for such quantities. The  reader is referred there for more details. Turning to the model in question, it is straightforward to apply the formulae derived in refs.~\cite{johnson:2024bue,Ahmed:2025lxe} to the case here. (It is strikingly simple since 
$u_0(x){=}{\widetilde\Gamma}^2/x^2$, $u_2(x){=}{-}1/4x^2$ and all other $u_{2g}(x){=}0$.) Using $u_0(x){=}{\widetilde\Gamma}^2/x^2$ in equation~(\ref{eq:totally-awesome}), and evaluating at $x=\mu$ gives:
\begin{equation}
    Q_1(\mu,E)=-\frac{E \sqrt{E_0}}{16 \left(E_0-E  \right)^{\frac{5}{2}} {\widetilde\Gamma}}\ ,
\end{equation}and since this comes with a single power of $\hbar$ in defining a $\rho(E)$ correction (see~(\ref{eq:rho-relation})), this defines $R_1(E)$ in
 equation~(\ref{eq:R-terms}), the first of the series of corrections. Expressions for $Q_2(x,E)$ and $Q_3(x,E)$ can be found in ref.~\cite{Ahmed:2025lxe}, and yield, (with factors $\hbar^3$ and $\hbar^5$ respectively) the remaining terms displayed in~(\ref{eq:R-terms}) and~(\ref{eq:extra-term}).

 The reader might wonder why one goes to all this trouble for this simple model. After all,  the exact spectral density is known explicitly  in terms of Bessel functions in this case, as shown in equation~(\ref{eq:exact-density-special}).  The simple answer is that the Bessel form actually obscures a lot of the  structures that played an important role in this paper, such as the organization in terms of inverse powers of $(E_0-E)^\frac12$. The polynomial structure in terms of powers of inverse $z^2$ is missed. Nevertheless, it is an amusing exercise to expand the exact $\rho(E)$ given in (\ref{eq:exact-density-special}) as a series in $\hbar$. One must average oscillatory trigonometric terms at every order, and then the result matches what one would get by expanding out the square roots in $E/E_0$, reversing the sign on $E,E_0$, and using $E_0{=}(-)\hbar^2\Gamma^2/\mu$. It works, but it hides a lot of what's elegant here.

\begin{widetext}
 \section{{\tt MATLAB} program for generating Wishart model}
 \label{app:matlab-program}
The accompanying file  is a simple  {\tt MATLAB} program for generating the illustrative data used in the body of the paper.  The first 40 lines does the random matrix model, the rest is concerned  with histogramming and plotting the data nicely. (A  {\tt parfor} loop over the samples (line 18) can be  used to speed things up considerably, but the {\tt lambda$\_$sorted$\{{\tt k}\}$} cell structure would have to be re-done.) The program is called: \begin{verbatim}
RMT_experiment_3_Wishart_scaling_Jan_2026_CVJ.m  \end{verbatim}

\begin{lstlisting}[frame=single,numbers=left,style=Matlab-editor]
%Experiment: Exploring Wishart model with  Marchencko-Pastur distribution
%and Bessel model in scaling limit - Clifford V. Johnson (January 9th 2026)
clear;
close all;
%% Setting up parameters and containers for data 
G=10; % short for Gamma
N=100; % the N in the text
n=N+G;
NS=10000; %number of samples in ensemble
Nlevels = 20; %number of energy levels to track statistics for
lambda=[];  % a bin for all the unsorted eigenvalues, empty at first.
lambda_sorted = cell(1, Nlevels); % bins for the sorted eigenvalues, empty.
for k = 1:Nlevels
    lambda_sorted{k} = [];
end
%% Time to sample now! 
tic %start measuring time
for i=1:NS  %distribute this over all the parallel cores

M=randn(n)+sqrt(-1)*randn(n); % make Gaussian random nxn complex matrix M

%M(102,:)=[]; %this is how you remove a row (reminder to self - CVJ)
%M(:,102)=[]; %this is how you remove a column (reminder to self - CVJ)

%So now remove a column G times from n=N+G square matrix
for j=0:G-1
 M(:,n-j)=[];
end

H=(M*M'); % this makes an nxn positive Hermitian matrix H
lambda=[lambda;eig(H)]; % update the bin of unsorted eigenvalues
sorted=sort(eig(H)); % sort the unsorted eigenvalues
% but we want to have sorted data too, so:
for k = 1:Nlevels
    lambda_sorted{k}(end+1,1) = sorted(k);
end % update the sorted bins eigenvalues

end %Done running over the ensemble. That's it. That's the random matrix model.
toc %report time taken

lambda=lambda/(4*N); % normalize eigenvalues to scale to finite cut and also match Marchenko-Pastur conventions.
%% Plot and histogram the data.
distribution=figure;

dx = 0.01; %picking a reaonable size bin.
edges = 0:dx:2.5; % setting the sample bin edges over a range

[freq ,x]=hist(lambda, edges); % histogramming the data

figure(distribution);
 
hold on
bar(x, freq/(NS*N*dx),'FaceColor',[0,0.5,1]) %plotting the histogrammed data, scaled by number of samples and size of matrix.
ym = G/N/2+1 - sqrt(G/N+1); %MP endpoint left
yp = G/N/2+1 + sqrt(G/N+1); %MP endpoint right
y=linspace(0,yp,5000); 
plot(y,sqrt((yp-y).*(y-ym))./(y*pi),'k', 'LineWidth' ,2) % plot MP distribution
axis([0 2.1 0 7])
xlabel('$\lambda$','Fontsize',30,'Interpreter','Latex'); ylabel('$\rho_0(\lambda)$','Fontsize',30,'rotation',0,'Interpreter','Latex');
hold off
%% Scaling in to the endpoint distribution part one.
scaled_distribution = figure;

e=4*lambda*(G/N+2)*N^2;%scaling up so that can effectively set hbar = 1 in exact formulae; see text)
dw = 5;
edgesE = 0:dw:1000; %setting the sample bin edges over a range
[freqEdge,w]=hist(e,0:dw:1000); %histogram the scaled data

Gt = 10; %Gamma tilde
tt =  linspace(0,1000,10000); % silly name for point grid for plotting the exact curves

exactdensity = (-0.2e1 * Gt * besselj(Gt, sqrt(tt)) .* besselj(Gt + 0.1e1, sqrt(tt)) + besselj(Gt, sqrt(tt)) .^ 2 .* sqrt(tt) + besselj(Gt + 0.1e1, sqrt(tt)) .^ 2 .* sqrt(tt)) .* tt .^ (-0.1e1 / 0.2e1) ./ 0.4e1;
classicaldensity = sqrt(-Gt ^ 2 + tt) / pi ./ tt / 0.2e1;

figure(scaled_distribution)
hold on
bar(w, freqEdge/(NS*dw),'FaceColor',[0.0,0.8,0])
plot(tt , exactdensity ,'-b', 'LineWidth' ,2)
plot(tt , classicaldensity ,'--b', 'LineWidth' ,2)
axis([0 700 0 0.012])
xlabel('$E$','Fontsize',25,'Interpreter','Latex'); ylabel('$\rho(E)$','Fontsize',25,'rotation',0,'Interpreter','Latex');
hold off
%% Scaling in to the endpoint distribution part two.
factor=4*(G/N+2)*N^2;%scaling up so that can set hbar =1 in exact formulae; see text)
factor = factor/4/N; %because I did not do it to the ordered bins before)
for k = 1:Nlevels
    lambda_sorted{k} = factor * lambda_sorted{k};
end
%% Plot and histogram the data.
dx = 5; %what was used in part 1 of scaling
distributions2=figure;

freqEdge = cell(1, Nlevels);
xEdge    = cell(1, Nlevels);

for k = 1:Nlevels
    xmax = max(lambda_sorted{k});
    edges = 0:dx:(1.1*xmax);  % hand-crafted to set edges of bins
    [freqEdge{k}, xEdge{k}] = hist(lambda_sorted{k}, edges);
end

figure(distributions2);
hold on
% how many ordered levels to overlay in the plot
Nplot = Nlevels;  
colours = lines(Nplot); % a color table of Nplot colours

% now plot them:
for k = 1:Nplot
    bar(xEdge{k}, freqEdge{k}/(NS*dx),'FaceColor', colours(k,:), 'EdgeColor', 'none');
end
tt=linspace(0,8000,30000); % silly name for point grid for plotting the exact curves
Gt = 10; %Gamma tilde
exactdensity = (-0.2e1 * Gt * besselj(Gt, sqrt(tt)) .* besselj(Gt + 0.1e1, sqrt(tt)) + besselj(Gt, sqrt(tt)) .^ 2 .* sqrt(tt) + besselj(Gt + 0.1e1, sqrt(tt)) .^ 2 .* sqrt(tt)) .* tt .^ (-0.1e1 / 0.2e1) ./ 0.4e1;
classicaldensity = sqrt(-Gt ^ 2 + tt) / pi ./ tt / 0.2e1;
plot(tt , exactdensity ,'-b', 'LineWidth' ,2)
plot(tt , classicaldensity,'--b', 'LineWidth' ,1)
axis([0 7000 0 0.012])
xlabel('$E$','Fontsize',25,'Interpreter','Latex'); ylabel('$\rho(E)$','Fontsize',25,'rotation',0,'Interpreter','Latex');
legend('$p(0;E)$','$p(1;E)$','$p(2;E)$','$p(3;E)$','$p(4;E)$','Fontsize',12,'Interpreter','Latex')
hold off

%% All done
load handel.mat; % Handel's Messiah lets me know program is done running.
sound(y);
\end{lstlisting}

\end{widetext}

\bibliographystyle{apsrev4-1}
\bibliography{Fredholm_super_JT_gravity1,Fredholm_super_JT_gravity2,extra_references,somerefs,Ref}

\begin{thebibliography}{79}%
\makeatletter
\providecommand \@ifxundefined [1]{%
 \@ifx{#1\undefined}
}%
\providecommand \@ifnum [1]{%
 \ifnum #1\expandafter \@firstoftwo
 \else \expandafter \@secondoftwo
 \fi
}%
\providecommand \@ifx [1]{%
 \ifx #1\expandafter \@firstoftwo
 \else \expandafter \@secondoftwo
 \fi
}%
\providecommand \natexlab [1]{#1}%
\providecommand \enquote  [1]{``#1''}%
\providecommand \bibnamefont  [1]{#1}%
\providecommand \bibfnamefont [1]{#1}%
\providecommand \citenamefont [1]{#1}%
\providecommand \href@noop [0]{\@secondoftwo}%
\providecommand \href [0]{\begingroup \@sanitize@url \@href}%
\providecommand \@href[1]{\@@startlink{#1}\@@href}%
\providecommand \@@href[1]{\endgroup#1\@@endlink}%
\providecommand \@sanitize@url [0]{\catcode `\\12\catcode `\$12\catcode `\&12\catcode `\#12\catcode `\^12\catcode `\_12\catcode `\%12\relax}%
\providecommand \@@startlink[1]{}%
\providecommand \@@endlink[0]{}%
\providecommand \url  [0]{\begingroup\@sanitize@url \@url }%
\providecommand \@url [1]{\endgroup\@href {#1}{\urlprefix }}%
\providecommand \urlprefix  [0]{URL }%
\providecommand \Eprint [0]{\href }%
\providecommand \doibase [0]{http://dx.doi.org/}%
\providecommand \selectlanguage [0]{\@gobble}%
\providecommand \bibinfo  [0]{\@secondoftwo}%
\providecommand \bibfield  [0]{\@secondoftwo}%
\providecommand \translation [1]{[#1]}%
\providecommand \BibitemOpen [0]{}%
\providecommand \bibitemStop [0]{}%
\providecommand \bibitemNoStop [0]{.\EOS\space}%
\providecommand \EOS [0]{\spacefactor3000\relax}%
\providecommand \BibitemShut  [1]{\csname bibitem#1\endcsname}%
\let\auto@bib@innerbib\@empty
\bibitem [{\citenamefont {Turiaci}\ and\ \citenamefont {Witten}(2023)}]{Turiaci:2023jfa}%
  \BibitemOpen
  \bibfield  {author} {\bibinfo {author} {\bibfnamefont {G.~J.}\ \bibnamefont {Turiaci}}\ and\ \bibinfo {author} {\bibfnamefont {E.}~\bibnamefont {Witten}},\ }\href {\doibase 10.1007/JHEP12(2023)003} {\bibfield  {journal} {\bibinfo  {journal} {JHEP}\ }\textbf {\bibinfo {volume} {12}},\ \bibinfo {pages} {003} (\bibinfo {year} {2023})},\ \Eprint {http://arxiv.org/abs/2305.19438} {arXiv:2305.19438 [hep-th]} \BibitemShut {NoStop}%
\bibitem [{\citenamefont {Johnson}(2024{\natexlab{a}})}]{Johnson:2023ofr}%
  \BibitemOpen
  \bibfield  {author} {\bibinfo {author} {\bibfnamefont {C.~V.}\ \bibnamefont {Johnson}},\ }\href {\doibase 10.1103/PhysRevD.110.106019} {\bibfield  {journal} {\bibinfo  {journal} {Phys. Rev. D}\ }\textbf {\bibinfo {volume} {110}},\ \bibinfo {pages} {106019} (\bibinfo {year} {2024}{\natexlab{a}})},\ \Eprint {http://arxiv.org/abs/2306.10139} {arXiv:2306.10139 [hep-th]} \BibitemShut {NoStop}%
\bibitem [{\citenamefont {Johnson}\ and\ \citenamefont {Usatyuk}(2024)}]{Johnson:2024tgg}%
  \BibitemOpen
  \bibfield  {author} {\bibinfo {author} {\bibfnamefont {C.~V.}\ \bibnamefont {Johnson}}\ and\ \bibinfo {author} {\bibfnamefont {M.}~\bibnamefont {Usatyuk}},\ }\href@noop {} {\  (\bibinfo {year} {2024})},\ \Eprint {http://arxiv.org/abs/2407.17583} {arXiv:2407.17583 [hep-th]} \BibitemShut {NoStop}%
\bibitem [{\citenamefont {Johnson}\ and\ \citenamefont {Kolanowski}(2025)}]{Johnson:2025oty}%
  \BibitemOpen
  \bibfield  {author} {\bibinfo {author} {\bibfnamefont {C.~V.}\ \bibnamefont {Johnson}}\ and\ \bibinfo {author} {\bibfnamefont {M.}~\bibnamefont {Kolanowski}},\ }\href@noop {} {\  (\bibinfo {year} {2025})},\ \Eprint {http://arxiv.org/abs/2507.07185} {arXiv:2507.07185 [hep-th]} \BibitemShut {NoStop}%
\bibitem [{\citenamefont {Iliesiu}\ and\ \citenamefont {Turiaci}(2021)}]{Iliesiu:2020qvm}%
  \BibitemOpen
  \bibfield  {author} {\bibinfo {author} {\bibfnamefont {L.~V.}\ \bibnamefont {Iliesiu}}\ and\ \bibinfo {author} {\bibfnamefont {G.~J.}\ \bibnamefont {Turiaci}},\ }\href {\doibase 10.1007/JHEP05(2021)145} {\bibfield  {journal} {\bibinfo  {journal} {JHEP}\ }\textbf {\bibinfo {volume} {05}},\ \bibinfo {pages} {145} (\bibinfo {year} {2021})},\ \Eprint {http://arxiv.org/abs/2003.02860} {arXiv:2003.02860 [hep-th]} \BibitemShut {NoStop}%
\bibitem [{\citenamefont {Heydeman}\ \emph {et~al.}(2022)\citenamefont {Heydeman}, \citenamefont {Iliesiu}, \citenamefont {Turiaci},\ and\ \citenamefont {Zhao}}]{Heydeman:2020hhw}%
  \BibitemOpen
  \bibfield  {author} {\bibinfo {author} {\bibfnamefont {M.}~\bibnamefont {Heydeman}}, \bibinfo {author} {\bibfnamefont {L.~V.}\ \bibnamefont {Iliesiu}}, \bibinfo {author} {\bibfnamefont {G.~J.}\ \bibnamefont {Turiaci}}, \ and\ \bibinfo {author} {\bibfnamefont {W.}~\bibnamefont {Zhao}},\ }\href {\doibase 10.1088/1751-8121/ac3be9} {\bibfield  {journal} {\bibinfo  {journal} {J. Phys. A}\ }\textbf {\bibinfo {volume} {55}},\ \bibinfo {pages} {014004} (\bibinfo {year} {2022})},\ \Eprint {http://arxiv.org/abs/2011.01953} {arXiv:2011.01953 [hep-th]} \BibitemShut {NoStop}%
\bibitem [{\citenamefont {Heydeman}\ \emph {et~al.}(2025)\citenamefont {Heydeman}, \citenamefont {Shi},\ and\ \citenamefont {Turiaci}}]{Heydeman:2025vcc}%
  \BibitemOpen
  \bibfield  {author} {\bibinfo {author} {\bibfnamefont {M.}~\bibnamefont {Heydeman}}, \bibinfo {author} {\bibfnamefont {X.}~\bibnamefont {Shi}}, \ and\ \bibinfo {author} {\bibfnamefont {G.~J.}\ \bibnamefont {Turiaci}},\ }\href@noop {} {\  (\bibinfo {year} {2025})},\ \Eprint {http://arxiv.org/abs/2504.20146} {arXiv:2504.20146 [hep-th]} \BibitemShut {NoStop}%
\bibitem [{\citenamefont {Chang}\ and\ \citenamefont {Lin}(2024)}]{Chang:2024zqi}%
  \BibitemOpen
  \bibfield  {author} {\bibinfo {author} {\bibfnamefont {C.-M.}\ \bibnamefont {Chang}}\ and\ \bibinfo {author} {\bibfnamefont {Y.-H.}\ \bibnamefont {Lin}},\ }\href@noop {} {\  (\bibinfo {year} {2024})},\ \Eprint {http://arxiv.org/abs/2402.10129} {arXiv:2402.10129 [hep-th]} \BibitemShut {NoStop}%
\bibitem [{\citenamefont {Lin}\ \emph {et~al.}(2023{\natexlab{a}})\citenamefont {Lin}, \citenamefont {Maldacena}, \citenamefont {Rozenberg},\ and\ \citenamefont {Shan}}]{Lin:2022rzw}%
  \BibitemOpen
  \bibfield  {author} {\bibinfo {author} {\bibfnamefont {H.~W.}\ \bibnamefont {Lin}}, \bibinfo {author} {\bibfnamefont {J.}~\bibnamefont {Maldacena}}, \bibinfo {author} {\bibfnamefont {L.}~\bibnamefont {Rozenberg}}, \ and\ \bibinfo {author} {\bibfnamefont {J.}~\bibnamefont {Shan}},\ }\href {\doibase 10.21468/SciPostPhys.14.6.150} {\bibfield  {journal} {\bibinfo  {journal} {SciPost Phys.}\ }\textbf {\bibinfo {volume} {14}},\ \bibinfo {pages} {150} (\bibinfo {year} {2023}{\natexlab{a}})},\ \Eprint {http://arxiv.org/abs/2207.00407} {arXiv:2207.00407 [hep-th]} \BibitemShut {NoStop}%
\bibitem [{\citenamefont {Lin}\ \emph {et~al.}(2023{\natexlab{b}})\citenamefont {Lin}, \citenamefont {Maldacena}, \citenamefont {Rozenberg},\ and\ \citenamefont {Shan}}]{Lin:2022zxd}%
  \BibitemOpen
  \bibfield  {author} {\bibinfo {author} {\bibfnamefont {H.~W.}\ \bibnamefont {Lin}}, \bibinfo {author} {\bibfnamefont {J.}~\bibnamefont {Maldacena}}, \bibinfo {author} {\bibfnamefont {L.}~\bibnamefont {Rozenberg}}, \ and\ \bibinfo {author} {\bibfnamefont {J.}~\bibnamefont {Shan}},\ }\href {\doibase 10.21468/SciPostPhys.14.5.128} {\bibfield  {journal} {\bibinfo  {journal} {SciPost Phys.}\ }\textbf {\bibinfo {volume} {14}},\ \bibinfo {pages} {128} (\bibinfo {year} {2023}{\natexlab{b}})},\ \Eprint {http://arxiv.org/abs/2207.00408} {arXiv:2207.00408 [hep-th]} \BibitemShut {NoStop}%
\bibitem [{\citenamefont {Chen}\ \emph {et~al.}(2025)\citenamefont {Chen}, \citenamefont {Lin},\ and\ \citenamefont {Shenker}}]{Chen:2024oqv}%
  \BibitemOpen
  \bibfield  {author} {\bibinfo {author} {\bibfnamefont {Y.}~\bibnamefont {Chen}}, \bibinfo {author} {\bibfnamefont {H.~W.}\ \bibnamefont {Lin}}, \ and\ \bibinfo {author} {\bibfnamefont {S.~H.}\ \bibnamefont {Shenker}},\ }\href {\doibase 10.21468/SciPostPhys.18.2.072} {\bibfield  {journal} {\bibinfo  {journal} {SciPost Phys.}\ }\textbf {\bibinfo {volume} {18}},\ \bibinfo {pages} {072} (\bibinfo {year} {2025})},\ \Eprint {http://arxiv.org/abs/2407.19387} {arXiv:2407.19387 [hep-th]} \BibitemShut {NoStop}%
\bibitem [{\citenamefont {Chang}\ \emph {et~al.}(2025)\citenamefont {Chang}, \citenamefont {Chen}, \citenamefont {Sia},\ and\ \citenamefont {Yang}}]{Chang:2024lxt}%
  \BibitemOpen
  \bibfield  {author} {\bibinfo {author} {\bibfnamefont {C.-M.}\ \bibnamefont {Chang}}, \bibinfo {author} {\bibfnamefont {Y.}~\bibnamefont {Chen}}, \bibinfo {author} {\bibfnamefont {B.~S.}\ \bibnamefont {Sia}}, \ and\ \bibinfo {author} {\bibfnamefont {Z.}~\bibnamefont {Yang}},\ }\href {\doibase 10.1007/JHEP08(2025)003} {\bibfield  {journal} {\bibinfo  {journal} {JHEP}\ }\textbf {\bibinfo {volume} {08}},\ \bibinfo {pages} {003} (\bibinfo {year} {2025})},\ \Eprint {http://arxiv.org/abs/2412.06902} {arXiv:2412.06902 [hep-th]} \BibitemShut {NoStop}%
\bibitem [{\citenamefont {Maldacena}\ \emph {et~al.}(2016{\natexlab{a}})\citenamefont {Maldacena}, \citenamefont {Shenker},\ and\ \citenamefont {Stanford}}]{Maldacena:2015waa}%
  \BibitemOpen
  \bibfield  {author} {\bibinfo {author} {\bibfnamefont {J.}~\bibnamefont {Maldacena}}, \bibinfo {author} {\bibfnamefont {S.~H.}\ \bibnamefont {Shenker}}, \ and\ \bibinfo {author} {\bibfnamefont {D.}~\bibnamefont {Stanford}},\ }\href {\doibase 10.1007/JHEP08(2016)106} {\bibfield  {journal} {\bibinfo  {journal} {JHEP}\ }\textbf {\bibinfo {volume} {08}},\ \bibinfo {pages} {106} (\bibinfo {year} {2016}{\natexlab{a}})},\ \Eprint {http://arxiv.org/abs/1503.01409} {arXiv:1503.01409 [hep-th]} \BibitemShut {NoStop}%
\bibitem [{\citenamefont {Mathur}\ and\ \citenamefont {Mehta}(2024)}]{Mathur:2024ify}%
  \BibitemOpen
  \bibfield  {author} {\bibinfo {author} {\bibfnamefont {S.~D.}\ \bibnamefont {Mathur}}\ and\ \bibinfo {author} {\bibfnamefont {M.}~\bibnamefont {Mehta}},\ }\href@noop {} {\  (\bibinfo {year} {2024})},\ \Eprint {http://arxiv.org/abs/2412.09495} {arXiv:2412.09495 [hep-th]} \BibitemShut {NoStop}%
\bibitem [{\citenamefont {Maldacena}\ and\ \citenamefont {Stanford}(2016)}]{Maldacena:2016hyu}%
  \BibitemOpen
  \bibfield  {author} {\bibinfo {author} {\bibfnamefont {J.}~\bibnamefont {Maldacena}}\ and\ \bibinfo {author} {\bibfnamefont {D.}~\bibnamefont {Stanford}},\ }\href {\doibase 10.1103/PhysRevD.94.106002} {\bibfield  {journal} {\bibinfo  {journal} {Phys. Rev.}\ }\textbf {\bibinfo {volume} {D94}},\ \bibinfo {pages} {106002} (\bibinfo {year} {2016})},\ \Eprint {http://arxiv.org/abs/1604.07818} {arXiv:1604.07818 [hep-th]} \BibitemShut {NoStop}%
\bibitem [{\citenamefont {Jensen}(2016)}]{Jensen:2016pah}%
  \BibitemOpen
  \bibfield  {author} {\bibinfo {author} {\bibfnamefont {K.}~\bibnamefont {Jensen}},\ }\href {\doibase 10.1103/PhysRevLett.117.111601} {\bibfield  {journal} {\bibinfo  {journal} {Phys. Rev. Lett.}\ }\textbf {\bibinfo {volume} {117}},\ \bibinfo {pages} {111601} (\bibinfo {year} {2016})},\ \Eprint {http://arxiv.org/abs/1605.06098} {arXiv:1605.06098 [hep-th]} \BibitemShut {NoStop}%
\bibitem [{\citenamefont {Maldacena}\ \emph {et~al.}(2016{\natexlab{b}})\citenamefont {Maldacena}, \citenamefont {Stanford},\ and\ \citenamefont {Yang}}]{Maldacena:2016upp}%
  \BibitemOpen
  \bibfield  {author} {\bibinfo {author} {\bibfnamefont {J.}~\bibnamefont {Maldacena}}, \bibinfo {author} {\bibfnamefont {D.}~\bibnamefont {Stanford}}, \ and\ \bibinfo {author} {\bibfnamefont {Z.}~\bibnamefont {Yang}},\ }\href {\doibase 10.1093/ptep/ptw124} {\bibfield  {journal} {\bibinfo  {journal} {PTEP}\ }\textbf {\bibinfo {volume} {2016}},\ \bibinfo {pages} {12C104} (\bibinfo {year} {2016}{\natexlab{b}})},\ \Eprint {http://arxiv.org/abs/1606.01857} {arXiv:1606.01857 [hep-th]} \BibitemShut {NoStop}%
\bibitem [{\citenamefont {Engels{\"o}y}\ \emph {et~al.}(2016)\citenamefont {Engels{\"o}y}, \citenamefont {Mertens},\ and\ \citenamefont {Verlinde}}]{Engelsoy:2016xyb}%
  \BibitemOpen
  \bibfield  {author} {\bibinfo {author} {\bibfnamefont {J.}~\bibnamefont {Engels{\"o}y}}, \bibinfo {author} {\bibfnamefont {T.~G.}\ \bibnamefont {Mertens}}, \ and\ \bibinfo {author} {\bibfnamefont {H.}~\bibnamefont {Verlinde}},\ }\href {\doibase 10.1007/JHEP07(2016)139} {\bibfield  {journal} {\bibinfo  {journal} {JHEP}\ }\textbf {\bibinfo {volume} {07}},\ \bibinfo {pages} {139} (\bibinfo {year} {2016})},\ \Eprint {http://arxiv.org/abs/1606.03438} {arXiv:1606.03438 [hep-th]} \BibitemShut {NoStop}%
\bibitem [{\citenamefont {Cotler}\ \emph {et~al.}(2017)\citenamefont {Cotler}, \citenamefont {Gur-Ari}, \citenamefont {Hanada}, \citenamefont {Polchinski}, \citenamefont {Saad}, \citenamefont {Shenker}, \citenamefont {Stanford}, \citenamefont {Streicher},\ and\ \citenamefont {Tezuka}}]{Cotler:2016fpe}%
  \BibitemOpen
  \bibfield  {author} {\bibinfo {author} {\bibfnamefont {J.~S.}\ \bibnamefont {Cotler}}, \bibinfo {author} {\bibfnamefont {G.}~\bibnamefont {Gur-Ari}}, \bibinfo {author} {\bibfnamefont {M.}~\bibnamefont {Hanada}}, \bibinfo {author} {\bibfnamefont {J.}~\bibnamefont {Polchinski}}, \bibinfo {author} {\bibfnamefont {P.}~\bibnamefont {Saad}}, \bibinfo {author} {\bibfnamefont {S.~H.}\ \bibnamefont {Shenker}}, \bibinfo {author} {\bibfnamefont {D.}~\bibnamefont {Stanford}}, \bibinfo {author} {\bibfnamefont {A.}~\bibnamefont {Streicher}}, \ and\ \bibinfo {author} {\bibfnamefont {M.}~\bibnamefont {Tezuka}},\ }\href {\doibase 10.1007/JHEP09(2018)002, 10.1007/JHEP05(2017)118} {\bibfield  {journal} {\bibinfo  {journal} {JHEP}\ }\textbf {\bibinfo {volume} {05}},\ \bibinfo {pages} {118} (\bibinfo {year} {2017})},\ \bibinfo {note} {[Erratum: JHEP09,002(2018)]},\ \Eprint {http://arxiv.org/abs/1611.04650} {arXiv:1611.04650 [hep-th]} \BibitemShut {NoStop}%
\bibitem [{\citenamefont {Preskill}\ \emph {et~al.}(1991)\citenamefont {Preskill}, \citenamefont {Schwarz}, \citenamefont {Shapere}, \citenamefont {Trivedi},\ and\ \citenamefont {Wilczek}}]{Preskill:1991tb}%
  \BibitemOpen
  \bibfield  {author} {\bibinfo {author} {\bibfnamefont {J.}~\bibnamefont {Preskill}}, \bibinfo {author} {\bibfnamefont {P.}~\bibnamefont {Schwarz}}, \bibinfo {author} {\bibfnamefont {A.~D.}\ \bibnamefont {Shapere}}, \bibinfo {author} {\bibfnamefont {S.}~\bibnamefont {Trivedi}}, \ and\ \bibinfo {author} {\bibfnamefont {F.}~\bibnamefont {Wilczek}},\ }\href {\doibase 10.1142/S0217732391002773} {\bibfield  {journal} {\bibinfo  {journal} {Mod. Phys. Lett. A}\ }\textbf {\bibinfo {volume} {6}},\ \bibinfo {pages} {2353} (\bibinfo {year} {1991})}\BibitemShut {NoStop}%
\bibitem [{\citenamefont {Turiaci}(2023)}]{Turiaci:2023wrh}%
  \BibitemOpen
  \bibfield  {author} {\bibinfo {author} {\bibfnamefont {G.~J.}\ \bibnamefont {Turiaci}},\ }\href@noop {} {\  (\bibinfo {year} {2023})},\ \Eprint {http://arxiv.org/abs/2307.10423} {arXiv:2307.10423 [hep-th]} \BibitemShut {NoStop}%
\bibitem [{\citenamefont {Saad}\ \emph {et~al.}(2019)\citenamefont {Saad}, \citenamefont {Shenker},\ and\ \citenamefont {Stanford}}]{Saad:2019lba}%
  \BibitemOpen
  \bibfield  {author} {\bibinfo {author} {\bibfnamefont {P.}~\bibnamefont {Saad}}, \bibinfo {author} {\bibfnamefont {S.~H.}\ \bibnamefont {Shenker}}, \ and\ \bibinfo {author} {\bibfnamefont {D.}~\bibnamefont {Stanford}},\ }\href@noop {} {\  (\bibinfo {year} {2019})},\ \Eprint {http://arxiv.org/abs/1903.11115} {arXiv:1903.11115 [hep-th]} \BibitemShut {NoStop}%
\bibitem [{\citenamefont {Rodrigues}\ \emph {et~al.}(pear)\citenamefont {Rodrigues}, \citenamefont {Schiappa},\ and\ \citenamefont {Saraswat}}]{joaoEtAl}%
  \BibitemOpen
  \bibfield  {author} {\bibinfo {author} {\bibfnamefont {J.}~\bibnamefont {Rodrigues}}, \bibinfo {author} {\bibfnamefont {R.}~\bibnamefont {Schiappa}}, \ and\ \bibinfo {author} {\bibfnamefont {K.}~\bibnamefont {Saraswat}},\ }\href@noop {} {\  (\bibinfo {year} {2026 (to appear)})}\BibitemShut {NoStop}%
\bibitem [{\citenamefont {Forrester}(2010)}]{ForresterBook}%
  \BibitemOpen
  \bibfield  {author} {\bibinfo {author} {\bibfnamefont {P.}~\bibnamefont {Forrester}},\ }\href {\doibase 10.1515/9781400835416} {\emph {\bibinfo {title} {Log-Gases and Random Matrices (LMS-34), {\rm Princeton University Press}}}}\ (\bibinfo {year} {2010})\BibitemShut {NoStop}%
\bibitem [{\citenamefont {Collins}(2005)}]{Collins:2005}%
  \BibitemOpen
  \bibfield  {author} {\bibinfo {author} {\bibfnamefont {B.}~\bibnamefont {Collins}},\ }\href {\doibase 10.1007/s00440-005-0428-5} {\bibfield  {journal} {\bibinfo  {journal} {Probab. Theory Relat. Fields}\ }\textbf {\bibinfo {volume} {133}},\ \bibinfo {pages} {315} (\bibinfo {year} {2005})},\ \Eprint {http://arxiv.org/abs/math/0406560} {arXiv:math/0406560 [math.PR]} \BibitemShut {NoStop}%
\bibitem [{\citenamefont {Johnson}(2024{\natexlab{b}})}]{johnson:2024bue}%
  \BibitemOpen
  \bibfield  {author} {\bibinfo {author} {\bibfnamefont {C.~V.}\ \bibnamefont {Johnson}},\ }\href {\doibase 10.1103/PhysRevD.110.066015} {\bibfield  {journal} {\bibinfo  {journal} {Phys. Rev. D}\ }\textbf {\bibinfo {volume} {110}},\ \bibinfo {pages} {066015} (\bibinfo {year} {2024}{\natexlab{b}})},\ \Eprint {http://arxiv.org/abs/2401.06220} {arXiv:2401.06220 [hep-th]} \BibitemShut {NoStop}%
\bibitem [{\citenamefont {Johnson}(2024{\natexlab{c}})}]{Johnson:2024fkm}%
  \BibitemOpen
  \bibfield  {author} {\bibinfo {author} {\bibfnamefont {C.~V.}\ \bibnamefont {Johnson}},\ }\href {\doibase 10.1103/PhysRevD.110.066016} {\bibfield  {journal} {\bibinfo  {journal} {Phys. Rev. D}\ }\textbf {\bibinfo {volume} {110}},\ \bibinfo {pages} {066016} (\bibinfo {year} {2024}{\natexlab{c}})},\ \Eprint {http://arxiv.org/abs/2401.08786} {arXiv:2401.08786 [hep-th]} \BibitemShut {NoStop}%
\bibitem [{\citenamefont {Lowenstein}(2024)}]{Lowenstein:2024gvz}%
  \BibitemOpen
  \bibfield  {author} {\bibinfo {author} {\bibfnamefont {A.}~\bibnamefont {Lowenstein}},\ }\href {\doibase 10.1007/JHEP07(2024)056} {\bibfield  {journal} {\bibinfo  {journal} {JHEP}\ }\textbf {\bibinfo {volume} {07}},\ \bibinfo {pages} {056} (\bibinfo {year} {2024})},\ \Eprint {http://arxiv.org/abs/2404.13175} {arXiv:2404.13175 [hep-th]} \BibitemShut {NoStop}%
\bibitem [{\citenamefont {Ahmed}\ \emph {et~al.}(2025)\citenamefont {Ahmed}, \citenamefont {Johnson},\ and\ \citenamefont {Saraswat}}]{Ahmed:2025lxe}%
  \BibitemOpen
  \bibfield  {author} {\bibinfo {author} {\bibfnamefont {W.}~\bibnamefont {Ahmed}}, \bibinfo {author} {\bibfnamefont {C.~V.}\ \bibnamefont {Johnson}}, \ and\ \bibinfo {author} {\bibfnamefont {K.}~\bibnamefont {Saraswat}},\ }\href@noop {} {\  (\bibinfo {year} {2025})},\ \Eprint {http://arxiv.org/abs/2507.18715} {arXiv:2507.18715 [hep-th]} \BibitemShut {NoStop}%
\bibitem [{\citenamefont {Eynard}\ and\ \citenamefont {Orantin}(2007{\natexlab{a}})}]{Eynard:2007kz}%
  \BibitemOpen
  \bibfield  {author} {\bibinfo {author} {\bibfnamefont {B.}~\bibnamefont {Eynard}}\ and\ \bibinfo {author} {\bibfnamefont {N.}~\bibnamefont {Orantin}},\ }\href {\doibase 10.4310/CNTP.2007.v1.n2.a4} {\bibfield  {journal} {\bibinfo  {journal} {Commun. Num. Theor. Phys.}\ }\textbf {\bibinfo {volume} {1}},\ \bibinfo {pages} {347} (\bibinfo {year} {2007}{\natexlab{a}})},\ \Eprint {http://arxiv.org/abs/math-ph/0702045} {arXiv:math-ph/0702045 [math-ph]} \BibitemShut {NoStop}%
\bibitem [{\citenamefont {Eynard}\ and\ \citenamefont {Orantin}(2007{\natexlab{b}})}]{Eynard:2007fi}%
  \BibitemOpen
  \bibfield  {author} {\bibinfo {author} {\bibfnamefont {B.}~\bibnamefont {Eynard}}\ and\ \bibinfo {author} {\bibfnamefont {N.}~\bibnamefont {Orantin}},\ }\href@noop {} {\  (\bibinfo {year} {2007}{\natexlab{b}})},\ \Eprint {http://arxiv.org/abs/0705.3600} {arXiv:0705.3600 [math-ph]} \BibitemShut {NoStop}%
\bibitem [{\citenamefont {Eynard}(2016)}]{Eynard:2016yaa}%
  \BibitemOpen
  \bibfield  {author} {\bibinfo {author} {\bibfnamefont {B.}~\bibnamefont {Eynard}},\ }\href {\doibase 10.1007/978-3-7643-8797-6} {\emph {\bibinfo {title} {{Counting Surfaces}}}},\ \bibinfo {series} {Progress in Mathematical Physics}, Vol.~\bibinfo {volume} {70}\ (\bibinfo  {publisher} {Springer},\ \bibinfo {year} {2016})\BibitemShut {NoStop}%
\bibitem [{\citenamefont {Do}(2008)}]{Do2008TouristGuide}%
  \BibitemOpen
  \bibfield  {author} {\bibinfo {author} {\bibfnamefont {N.}~\bibnamefont {Do}},\ }\href {users.monash.edu.au} {\bibfield  {journal} {\bibinfo  {journal} {Gazette of the Australian Mathematical Society}\ }\textbf {\bibinfo {volume} {35}},\ \bibinfo {pages} {103} (\bibinfo {year} {2008})}\BibitemShut {NoStop}%
\bibitem [{\citenamefont {Witten}(1990)}]{Witten:1989ig}%
  \BibitemOpen
  \bibfield  {author} {\bibinfo {author} {\bibfnamefont {E.}~\bibnamefont {Witten}},\ }\href {\doibase 10.1016/0550-3213(90)90449-N} {\bibfield  {journal} {\bibinfo  {journal} {Nucl. Phys. B}\ }\textbf {\bibinfo {volume} {340}},\ \bibinfo {pages} {281} (\bibinfo {year} {1990})}\BibitemShut {NoStop}%
\bibitem [{\citenamefont {Witten}(1991)}]{Witten:1990hr}%
  \BibitemOpen
  \bibfield  {author} {\bibinfo {author} {\bibfnamefont {E.}~\bibnamefont {Witten}},\ }\href {\doibase 10.4310/SDG.1990.v1.n1.a5} {\bibfield  {journal} {\bibinfo  {journal} {Surveys Diff. Geom.}\ }\textbf {\bibinfo {volume} {1}},\ \bibinfo {pages} {243} (\bibinfo {year} {1991})}\BibitemShut {NoStop}%
\bibitem [{\citenamefont {Kontsevich}(1992)}]{Kontsevich:1992ti}%
  \BibitemOpen
  \bibfield  {author} {\bibinfo {author} {\bibfnamefont {M.}~\bibnamefont {Kontsevich}},\ }\href {\doibase 10.1007/BF02099526} {\bibfield  {journal} {\bibinfo  {journal} {Commun. Math. Phys.}\ }\textbf {\bibinfo {volume} {147}},\ \bibinfo {pages} {1} (\bibinfo {year} {1992})}\BibitemShut {NoStop}%
\bibitem [{\citenamefont {Norbury}(2023)}]{Norbury:2017eih}%
  \BibitemOpen
  \bibfield  {author} {\bibinfo {author} {\bibfnamefont {P.}~\bibnamefont {Norbury}},\ }\href {\doibase 10.2140/gt.2023.27.2695} {\bibfield  {journal} {\bibinfo  {journal} {Geom. Topol.}\ }\textbf {\bibinfo {volume} {27}},\ \bibinfo {pages} {2695} (\bibinfo {year} {2023})},\ \Eprint {http://arxiv.org/abs/1712.03662} {arXiv:1712.03662 [math.AG]} \BibitemShut {NoStop}%
\bibitem [{\citenamefont {Guo}\ \emph {et~al.}(2025)\citenamefont {Guo}, \citenamefont {Norbury}, \citenamefont {Yang},\ and\ \citenamefont {Zagier}}]{guo2025combinatoricslargegenusasymptotics}%
  \BibitemOpen
  \bibfield  {author} {\bibinfo {author} {\bibfnamefont {J.}~\bibnamefont {Guo}}, \bibinfo {author} {\bibfnamefont {P.}~\bibnamefont {Norbury}}, \bibinfo {author} {\bibfnamefont {D.}~\bibnamefont {Yang}}, \ and\ \bibinfo {author} {\bibfnamefont {D.}~\bibnamefont {Zagier}},\ }\href {https://arxiv.org/abs/2412.20388} {\  (\bibinfo {year} {2025})},\ \Eprint {http://arxiv.org/abs/2412.20388} {arXiv:2412.20388 [math-ph]} \BibitemShut {NoStop}%
\bibitem [{\citenamefont {Brezin}\ and\ \citenamefont {Gross}(1980)}]{Brezin:1980rk}%
  \BibitemOpen
  \bibfield  {author} {\bibinfo {author} {\bibfnamefont {E.}~\bibnamefont {Brezin}}\ and\ \bibinfo {author} {\bibfnamefont {D.~J.}\ \bibnamefont {Gross}},\ }\href {\doibase 10.1016/0370-2693(80)90562-6} {\bibfield  {journal} {\bibinfo  {journal} {Phys. Lett. B}\ }\textbf {\bibinfo {volume} {97}},\ \bibinfo {pages} {120} (\bibinfo {year} {1980})}\BibitemShut {NoStop}%
\bibitem [{\citenamefont {Gross}\ and\ \citenamefont {Witten}(1980)}]{Gross:1980he}%
  \BibitemOpen
  \bibfield  {author} {\bibinfo {author} {\bibfnamefont {D.~J.}\ \bibnamefont {Gross}}\ and\ \bibinfo {author} {\bibfnamefont {E.}~\bibnamefont {Witten}},\ }\href {\doibase 10.1103/PhysRevD.21.446} {\bibfield  {journal} {\bibinfo  {journal} {Phys. Rev. D}\ }\textbf {\bibinfo {volume} {21}},\ \bibinfo {pages} {446} (\bibinfo {year} {1980})}\BibitemShut {NoStop}%
\bibitem [{\citenamefont {Gross}\ and\ \citenamefont {Newman}(1991)}]{Gross:1991aj}%
  \BibitemOpen
  \bibfield  {author} {\bibinfo {author} {\bibfnamefont {D.~J.}\ \bibnamefont {Gross}}\ and\ \bibinfo {author} {\bibfnamefont {M.~J.}\ \bibnamefont {Newman}},\ }\href {\doibase 10.1016/0370-2693(91)91042-T} {\bibfield  {journal} {\bibinfo  {journal} {Phys. Lett. B}\ }\textbf {\bibinfo {volume} {266}},\ \bibinfo {pages} {291} (\bibinfo {year} {1991})}\BibitemShut {NoStop}%
\bibitem [{\citenamefont {Dalley}\ \emph {et~al.}(1992{\natexlab{a}})\citenamefont {Dalley}, \citenamefont {Johnson}, \citenamefont {Morris},\ and\ \citenamefont {Watterstam}}]{Dalley:1992br}%
  \BibitemOpen
  \bibfield  {author} {\bibinfo {author} {\bibfnamefont {S.}~\bibnamefont {Dalley}}, \bibinfo {author} {\bibfnamefont {C.~V.}\ \bibnamefont {Johnson}}, \bibinfo {author} {\bibfnamefont {T.~R.}\ \bibnamefont {Morris}}, \ and\ \bibinfo {author} {\bibfnamefont {A.}~\bibnamefont {Watterstam}},\ }\href@noop {} {\bibfield  {journal} {\bibinfo  {journal} {Mod. Phys. Lett.}\ }\textbf {\bibinfo {volume} {A7}},\ \bibinfo {pages} {2753} (\bibinfo {year} {1992}{\natexlab{a}})},\ \Eprint {http://arxiv.org/abs/hep-th/9206060} {hep-th/9206060} \BibitemShut {NoStop}%
\bibitem [{\citenamefont {Chidambaram}\ \emph {et~al.}(2025)\citenamefont {Chidambaram}, \citenamefont {Garcia-Failde},\ and\ \citenamefont {Giacchetto}}]{Chidambaram:2022cqc}%
  \BibitemOpen
  \bibfield  {author} {\bibinfo {author} {\bibfnamefont {N.~K.}\ \bibnamefont {Chidambaram}}, \bibinfo {author} {\bibfnamefont {E.}~\bibnamefont {Garcia-Failde}}, \ and\ \bibinfo {author} {\bibfnamefont {A.}~\bibnamefont {Giacchetto}},\ }\href {\doibase 10.1007/s00222-025-01351-y} {\bibfield  {journal} {\bibinfo  {journal} {Invent. Math.}\ }\textbf {\bibinfo {volume} {241}},\ \bibinfo {pages} {929} (\bibinfo {year} {2025})},\ \Eprint {http://arxiv.org/abs/2205.15621} {arXiv:2205.15621 [math.AG]} \BibitemShut {NoStop}%
\bibitem [{\citenamefont {Iwaki}\ \emph {et~al.}(2018)\citenamefont {Iwaki}, \citenamefont {Koike},\ and\ \citenamefont {Takei}}]{iwaki2018voroscoefficientshypergeometricdifferential}%
  \BibitemOpen
  \bibfield  {author} {\bibinfo {author} {\bibfnamefont {K.}~\bibnamefont {Iwaki}}, \bibinfo {author} {\bibfnamefont {T.}~\bibnamefont {Koike}}, \ and\ \bibinfo {author} {\bibfnamefont {Y.}~\bibnamefont {Takei}},\ }\href {https://arxiv.org/abs/1810.02946} {\  (\bibinfo {year} {2018})},\ \Eprint {http://arxiv.org/abs/1810.02946} {arXiv:1810.02946 [math.CA]} \BibitemShut {NoStop}%
\bibitem [{\citenamefont {Bouchard}\ \emph {et~al.}(2025)\citenamefont {Bouchard}, \citenamefont {Chidambaram}, \citenamefont {Giacchetto},\ and\ \citenamefont {Shadrin}}]{bouchard2025thetaclassesgeneralizedtopological}%
  \BibitemOpen
  \bibfield  {author} {\bibinfo {author} {\bibfnamefont {V.}~\bibnamefont {Bouchard}}, \bibinfo {author} {\bibfnamefont {N.~K.}\ \bibnamefont {Chidambaram}}, \bibinfo {author} {\bibfnamefont {A.}~\bibnamefont {Giacchetto}}, \ and\ \bibinfo {author} {\bibfnamefont {S.}~\bibnamefont {Shadrin}},\ }\href {https://arxiv.org/abs/2505.11291} {\  (\bibinfo {year} {2025})},\ \Eprint {http://arxiv.org/abs/2505.11291} {arXiv:2505.11291 [math.AG]} \BibitemShut {NoStop}%
\bibitem [{\citenamefont {Mironov}\ \emph {et~al.}(1996)\citenamefont {Mironov}, \citenamefont {Morozov},\ and\ \citenamefont {Semenoff}}]{Mironov:1994mv}%
  \BibitemOpen
  \bibfield  {author} {\bibinfo {author} {\bibfnamefont {A.}~\bibnamefont {Mironov}}, \bibinfo {author} {\bibfnamefont {A.}~\bibnamefont {Morozov}}, \ and\ \bibinfo {author} {\bibfnamefont {G.~W.}\ \bibnamefont {Semenoff}},\ }\href {\doibase 10.1142/S0217751X96002339} {\bibfield  {journal} {\bibinfo  {journal} {Int. J. Mod. Phys. A}\ }\textbf {\bibinfo {volume} {11}},\ \bibinfo {pages} {5031} (\bibinfo {year} {1996})},\ \Eprint {http://arxiv.org/abs/hep-th/9404005} {arXiv:hep-th/9404005} \BibitemShut {NoStop}%
\bibitem [{\citenamefont {Alexandrov}(2018)}]{Alexandrov:2016kjl}%
  \BibitemOpen
  \bibfield  {author} {\bibinfo {author} {\bibfnamefont {A.}~\bibnamefont {Alexandrov}},\ }\href {\doibase 10.4310/ATMP.2018.v22.n6.a1} {\bibfield  {journal} {\bibinfo  {journal} {Adv. Theor. Math. Phys.}\ }\textbf {\bibinfo {volume} {22}},\ \bibinfo {pages} {1347} (\bibinfo {year} {2018})},\ \Eprint {http://arxiv.org/abs/1608.01627} {arXiv:1608.01627 [math-ph]} \BibitemShut {NoStop}%
\bibitem [{\citenamefont {Dubrovin}\ \emph {et~al.}(2021)\citenamefont {Dubrovin}, \citenamefont {Yang},\ and\ \citenamefont {Zagier}}]{Dubrovin:2018cho}%
  \BibitemOpen
  \bibfield  {author} {\bibinfo {author} {\bibfnamefont {B.}~\bibnamefont {Dubrovin}}, \bibinfo {author} {\bibfnamefont {D.}~\bibnamefont {Yang}}, \ and\ \bibinfo {author} {\bibfnamefont {D.}~\bibnamefont {Zagier}},\ }\href {\doibase 10.1007/s00029-021-00620-x} {\bibfield  {journal} {\bibinfo  {journal} {Selecta Math.}\ }\textbf {\bibinfo {volume} {27}},\ \bibinfo {pages} {12} (\bibinfo {year} {2021})},\ \Eprint {http://arxiv.org/abs/1812.08488} {arXiv:1812.08488 [math-ph]} \BibitemShut {NoStop}%
\bibitem [{\citenamefont {Alexandrov}(2023)}]{Alexandrov:2021etm}%
  \BibitemOpen
  \bibfield  {author} {\bibinfo {author} {\bibfnamefont {A.}~\bibnamefont {Alexandrov}},\ }\href {\doibase 10.1016/j.aim.2022.108809} {\bibfield  {journal} {\bibinfo  {journal} {Adv. Math.}\ }\textbf {\bibinfo {volume} {412}},\ \bibinfo {pages} {108809} (\bibinfo {year} {2023})},\ \Eprint {http://arxiv.org/abs/2103.17117} {arXiv:2103.17117 [math-ph]} \BibitemShut {NoStop}%
\bibitem [{\citenamefont {Mertens}\ and\ \citenamefont {Turiaci}(2021)}]{Mertens:2020hbs}%
  \BibitemOpen
  \bibfield  {author} {\bibinfo {author} {\bibfnamefont {T.~G.}\ \bibnamefont {Mertens}}\ and\ \bibinfo {author} {\bibfnamefont {G.~J.}\ \bibnamefont {Turiaci}},\ }\href {\doibase 10.1007/JHEP01(2021)073} {\bibfield  {journal} {\bibinfo  {journal} {JHEP}\ }\textbf {\bibinfo {volume} {01}},\ \bibinfo {pages} {073} (\bibinfo {year} {2021})},\ \Eprint {http://arxiv.org/abs/2006.07072} {arXiv:2006.07072 [hep-th]} \BibitemShut {NoStop}%
\bibitem [{\citenamefont {Wishart}(1928)}]{10.2307/2331939}%
  \BibitemOpen
  \bibfield  {author} {\bibinfo {author} {\bibfnamefont {J.}~\bibnamefont {Wishart}},\ }\href {http://www.jstor.org/stable/2331939} {\bibfield  {journal} {\bibinfo  {journal} {Biometrika}\ }\textbf {\bibinfo {volume} {20A}},\ \bibinfo {pages} {32} (\bibinfo {year} {1928})}\BibitemShut {NoStop}%
\bibitem [{\citenamefont {Morris}()}]{Morris:1990bw}%
  \BibitemOpen
  \bibfield  {author} {\bibinfo {author} {\bibfnamefont {T.~R.}\ \bibnamefont {Morris}},\ }\href@noop {} {\ }\bibinfo {note} {FERMILAB-PUB-90-136-T}\BibitemShut {NoStop}%
\bibitem [{\citenamefont {Morris}(1991)}]{Morris:1991cq}%
  \BibitemOpen
  \bibfield  {author} {\bibinfo {author} {\bibfnamefont {T.~R.}\ \bibnamefont {Morris}},\ }\href@noop {} {\bibfield  {journal} {\bibinfo  {journal} {Nucl. Phys.}\ }\textbf {\bibinfo {volume} {B356}},\ \bibinfo {pages} {703} (\bibinfo {year} {1991})}\BibitemShut {NoStop}%
\bibitem [{\citenamefont {Dalley}\ \emph {et~al.}(1992{\natexlab{b}})\citenamefont {Dalley}, \citenamefont {Johnson},\ and\ \citenamefont {Morris}}]{Dalley:1991qg}%
  \BibitemOpen
  \bibfield  {author} {\bibinfo {author} {\bibfnamefont {S.}~\bibnamefont {Dalley}}, \bibinfo {author} {\bibfnamefont {C.~V.}\ \bibnamefont {Johnson}}, \ and\ \bibinfo {author} {\bibfnamefont {T.}~\bibnamefont {Morris}},\ }\href@noop {} {\bibfield  {journal} {\bibinfo  {journal} {Nucl. Phys.}\ }\textbf {\bibinfo {volume} {B368}},\ \bibinfo {pages} {625} (\bibinfo {year} {1992}{\natexlab{b}})}\BibitemShut {NoStop}%
\bibitem [{\citenamefont {Anderson}\ \emph {et~al.}(1991)\citenamefont {Anderson}, \citenamefont {Myers},\ and\ \citenamefont {Periwal}}]{Anderson:1991ku}%
  \BibitemOpen
  \bibfield  {author} {\bibinfo {author} {\bibfnamefont {A.}~\bibnamefont {Anderson}}, \bibinfo {author} {\bibfnamefont {R.~C.}\ \bibnamefont {Myers}}, \ and\ \bibinfo {author} {\bibfnamefont {V.}~\bibnamefont {Periwal}},\ }\href {\doibase 10.1016/0550-3213(91)90411-P} {\bibfield  {journal} {\bibinfo  {journal} {Nucl. Phys. B}\ }\textbf {\bibinfo {volume} {360}},\ \bibinfo {pages} {463} (\bibinfo {year} {1991})}\BibitemShut {NoStop}%
\bibitem [{\citenamefont {Myers}\ and\ \citenamefont {Periwal}(1993)}]{Myers:1991akt}%
  \BibitemOpen
  \bibfield  {author} {\bibinfo {author} {\bibfnamefont {R.~C.}\ \bibnamefont {Myers}}\ and\ \bibinfo {author} {\bibfnamefont {V.}~\bibnamefont {Periwal}},\ }\href {\doibase 10.1016/0550-3213(93)90496-C} {\bibfield  {journal} {\bibinfo  {journal} {Nucl. Phys. B}\ }\textbf {\bibinfo {volume} {390}},\ \bibinfo {pages} {716} (\bibinfo {year} {1993})},\ \Eprint {http://arxiv.org/abs/hep-th/9112037} {arXiv:hep-th/9112037} \BibitemShut {NoStop}%
\bibitem [{\citenamefont {Lafrance}\ and\ \citenamefont {Myers}(1994)}]{Lafrance:1993wy}%
  \BibitemOpen
  \bibfield  {author} {\bibinfo {author} {\bibfnamefont {R.}~\bibnamefont {Lafrance}}\ and\ \bibinfo {author} {\bibfnamefont {R.~C.}\ \bibnamefont {Myers}},\ }\href {\doibase 10.1142/S0217732394000113} {\bibfield  {journal} {\bibinfo  {journal} {Mod. Phys. Lett. A}\ }\textbf {\bibinfo {volume} {9}},\ \bibinfo {pages} {101} (\bibinfo {year} {1994})},\ \Eprint {http://arxiv.org/abs/hep-th/9308113} {arXiv:hep-th/9308113} \BibitemShut {NoStop}%
\bibitem [{\citenamefont {Stanford}\ and\ \citenamefont {Witten}(2020)}]{Stanford:2019vob}%
  \BibitemOpen
  \bibfield  {author} {\bibinfo {author} {\bibfnamefont {D.}~\bibnamefont {Stanford}}\ and\ \bibinfo {author} {\bibfnamefont {E.}~\bibnamefont {Witten}},\ }\href {\doibase 10.4310/ATMP.2020.v24.n6.a4} {\bibfield  {journal} {\bibinfo  {journal} {Adv. Theor. Math. Phys.}\ }\textbf {\bibinfo {volume} {24}},\ \bibinfo {pages} {1475} (\bibinfo {year} {2020})},\ \Eprint {http://arxiv.org/abs/1907.03363} {arXiv:1907.03363 [hep-th]} \BibitemShut {NoStop}%
\bibitem [{\citenamefont {Johnson}(2021{\natexlab{a}})}]{Johnson:2020heh}%
  \BibitemOpen
  \bibfield  {author} {\bibinfo {author} {\bibfnamefont {C.~V.}\ \bibnamefont {Johnson}},\ }\href {\doibase 10.1103/PhysRevD.103.046012} {\bibfield  {journal} {\bibinfo  {journal} {Phys. Rev. D}\ }\textbf {\bibinfo {volume} {103}},\ \bibinfo {pages} {046012} (\bibinfo {year} {2021}{\natexlab{a}})},\ \Eprint {http://arxiv.org/abs/2005.01893} {arXiv:2005.01893 [hep-th]} \BibitemShut {NoStop}%
\bibitem [{\citenamefont {Johnson}(2021{\natexlab{b}})}]{Johnson:2020exp}%
  \BibitemOpen
  \bibfield  {author} {\bibinfo {author} {\bibfnamefont {C.~V.}\ \bibnamefont {Johnson}},\ }\href {\doibase 10.1103/PhysRevD.103.046013} {\bibfield  {journal} {\bibinfo  {journal} {Phys. Rev. D}\ }\textbf {\bibinfo {volume} {103}},\ \bibinfo {pages} {046013} (\bibinfo {year} {2021}{\natexlab{b}})},\ \Eprint {http://arxiv.org/abs/2006.10959} {arXiv:2006.10959 [hep-th]} \BibitemShut {NoStop}%
\bibitem [{\citenamefont {Bessis}\ \emph {et~al.}(1980)\citenamefont {Bessis}, \citenamefont {Itzykson},\ and\ \citenamefont {Zuber}}]{Bessis:1980ss}%
  \BibitemOpen
  \bibfield  {author} {\bibinfo {author} {\bibfnamefont {D.}~\bibnamefont {Bessis}}, \bibinfo {author} {\bibfnamefont {C.}~\bibnamefont {Itzykson}}, \ and\ \bibinfo {author} {\bibfnamefont {J.~B.}\ \bibnamefont {Zuber}},\ }\href@noop {} {\bibfield  {journal} {\bibinfo  {journal} {Adv. Appl. Math.}\ }\textbf {\bibinfo {volume} {1}},\ \bibinfo {pages} {109} (\bibinfo {year} {1980})}\BibitemShut {NoStop}%
\bibitem [{\citenamefont {Gross}\ and\ \citenamefont {Migdal}(1990)}]{Gross:1990aw}%
  \BibitemOpen
  \bibfield  {author} {\bibinfo {author} {\bibfnamefont {D.~J.}\ \bibnamefont {Gross}}\ and\ \bibinfo {author} {\bibfnamefont {A.~A.}\ \bibnamefont {Migdal}},\ }\href@noop {} {\bibfield  {journal} {\bibinfo  {journal} {Nucl. Phys.}\ }\textbf {\bibinfo {volume} {B340}},\ \bibinfo {pages} {333} (\bibinfo {year} {1990})}\BibitemShut {NoStop}%
\bibitem [{\citenamefont {Banks}\ \emph {et~al.}(1990)\citenamefont {Banks}, \citenamefont {Douglas}, \citenamefont {Seiberg},\ and\ \citenamefont {Shenker}}]{Banks:1990df}%
  \BibitemOpen
  \bibfield  {author} {\bibinfo {author} {\bibfnamefont {T.}~\bibnamefont {Banks}}, \bibinfo {author} {\bibfnamefont {M.~R.}\ \bibnamefont {Douglas}}, \bibinfo {author} {\bibfnamefont {N.}~\bibnamefont {Seiberg}}, \ and\ \bibinfo {author} {\bibfnamefont {S.~H.}\ \bibnamefont {Shenker}},\ }\href@noop {} {\bibfield  {journal} {\bibinfo  {journal} {Phys. Lett.}\ }\textbf {\bibinfo {volume} {B238}},\ \bibinfo {pages} {279} (\bibinfo {year} {1990})}\BibitemShut {NoStop}%
\bibitem [{\citenamefont {Gel'fand}\ and\ \citenamefont {Dikii}(1975)}]{Gelfand:1975rn}%
  \BibitemOpen
  \bibfield  {author} {\bibinfo {author} {\bibfnamefont {I.~M.}\ \bibnamefont {Gel'fand}}\ and\ \bibinfo {author} {\bibfnamefont {L.~A.}\ \bibnamefont {Dikii}},\ }\href@noop {} {\bibfield  {journal} {\bibinfo  {journal} {Russ. Math. Surveys}\ }\textbf {\bibinfo {volume} {30}},\ \bibinfo {pages} {77} (\bibinfo {year} {1975})}\BibitemShut {NoStop}%
\bibitem [{\citenamefont {Gel'fand}\ and\ \citenamefont {Dikii}(1976)}]{Gelfand:1976B}%
  \BibitemOpen
  \bibfield  {author} {\bibinfo {author} {\bibfnamefont {I.~M.}\ \bibnamefont {Gel'fand}}\ and\ \bibinfo {author} {\bibfnamefont {L.~A.}\ \bibnamefont {Dikii}},\ }\href@noop {} {\bibfield  {journal} {\bibinfo  {journal} {Funct.\ Anal.\ Appl.}\ }\textbf {\bibinfo {volume} {11}},\ \bibinfo {pages} {93} (\bibinfo {year} {1976})}\BibitemShut {NoStop}%
\bibitem [{\citenamefont {Carlisle}\ \emph {et~al.}(2008)\citenamefont {Carlisle}, \citenamefont {Johnson},\ and\ \citenamefont {Pennington}}]{Carlisle:2005wa}%
  \BibitemOpen
  \bibfield  {author} {\bibinfo {author} {\bibfnamefont {J.~E.}\ \bibnamefont {Carlisle}}, \bibinfo {author} {\bibfnamefont {C.~V.}\ \bibnamefont {Johnson}}, \ and\ \bibinfo {author} {\bibfnamefont {J.~S.}\ \bibnamefont {Pennington}},\ }\href {\doibase 10.1088/1751-8113/41/8/085401} {\bibfield  {journal} {\bibinfo  {journal} {J. Phys.}\ }\textbf {\bibinfo {volume} {A41}},\ \bibinfo {pages} {085401} (\bibinfo {year} {2008})},\ \Eprint {http://arxiv.org/abs/hep-th/0511002} {arXiv:hep-th/0511002 [hep-th]} \BibitemShut {NoStop}%
\bibitem [{\citenamefont {Nagao}\ and\ \citenamefont {Slevin}(1993)}]{doi:10.1063/1.530157}%
  \BibitemOpen
  \bibfield  {author} {\bibinfo {author} {\bibfnamefont {T.}~\bibnamefont {Nagao}}\ and\ \bibinfo {author} {\bibfnamefont {K.}~\bibnamefont {Slevin}},\ }\href {\doibase 10.1063/1.530157} {\bibfield  {journal} {\bibinfo  {journal} {Journal of Mathematical Physics}\ }\textbf {\bibinfo {volume} {34}},\ \bibinfo {pages} {2075} (\bibinfo {year} {1993})}\BibitemShut {NoStop}%
\bibitem [{\citenamefont {Morris}(1992)}]{Morris:1992zr}%
  \BibitemOpen
  \bibfield  {author} {\bibinfo {author} {\bibfnamefont {T.~R.}\ \bibnamefont {Morris}},\ }\href@noop {} {\bibfield  {journal} {\bibinfo  {journal} {Class. Quant. Grav.}\ }\textbf {\bibinfo {volume} {9}},\ \bibinfo {pages} {1873} (\bibinfo {year} {1992})}\BibitemShut {NoStop}%
\bibitem [{\citenamefont {Ince}(2012)}]{InceBook}%
  \BibitemOpen
  \bibfield  {author} {\bibinfo {author} {\bibfnamefont {E.~L.}\ \bibnamefont {Ince}},\ }\href@noop {} {\emph {\bibinfo {title} {Ordinary Differential Equations}}},\ \bibinfo {edition} {reprint}\ ed.,\ Dover Books on Mathematics\ (\bibinfo  {publisher} {Dover Publications},\ \bibinfo {address} {Mineola, NY, USA},\ \bibinfo {year} {2012})\BibitemShut {NoStop}%
\bibitem [{\citenamefont {Deutsch}(1991)}]{Deutsch:1991msp}%
  \BibitemOpen
  \bibfield  {author} {\bibinfo {author} {\bibfnamefont {J.~M.}\ \bibnamefont {Deutsch}},\ }\href {\doibase 10.1103/PhysRevA.43.2046} {\bibfield  {journal} {\bibinfo  {journal} {Phys. Rev. A}\ }\textbf {\bibinfo {volume} {43}},\ \bibinfo {pages} {2046} (\bibinfo {year} {1991})}\BibitemShut {NoStop}%
\bibitem [{\citenamefont {Srednicki}(1994)}]{Srednicki:1994mfb}%
  \BibitemOpen
  \bibfield  {author} {\bibinfo {author} {\bibfnamefont {M.}~\bibnamefont {Srednicki}},\ }\href {\doibase 10.1103/PhysRevE.50.888} {\bibfield  {journal} {\bibinfo  {journal} {Phys. Rev. E}\ }\textbf {\bibinfo {volume} {50}} (\bibinfo {year} {1994}),\ 10.1103/PhysRevE.50.888},\ \Eprint {http://arxiv.org/abs/cond-mat/9403051} {arXiv:cond-mat/9403051} \BibitemShut {NoStop}%
\bibitem [{\citenamefont {Rigol}\ \emph {et~al.}(2008)\citenamefont {Rigol}, \citenamefont {Dunjko},\ and\ \citenamefont {Olshanii}}]{Rigol:2007juv}%
  \BibitemOpen
  \bibfield  {author} {\bibinfo {author} {\bibfnamefont {M.}~\bibnamefont {Rigol}}, \bibinfo {author} {\bibfnamefont {V.}~\bibnamefont {Dunjko}}, \ and\ \bibinfo {author} {\bibfnamefont {M.}~\bibnamefont {Olshanii}},\ }\href {\doibase 10.1038/nature06838} {\bibfield  {journal} {\bibinfo  {journal} {Nature}\ }\textbf {\bibinfo {volume} {452}},\ \bibinfo {pages} {854} (\bibinfo {year} {2008})},\ \Eprint {http://arxiv.org/abs/0708.1324} {arXiv:0708.1324 [cond-mat.stat-mech]} \BibitemShut {NoStop}%
\bibitem [{\citenamefont {D'Alessio}\ \emph {et~al.}(2016)\citenamefont {D'Alessio}, \citenamefont {Kafri}, \citenamefont {Polkovnikov},\ and\ \citenamefont {Rigol}}]{DAlessio:2015qtq}%
  \BibitemOpen
  \bibfield  {author} {\bibinfo {author} {\bibfnamefont {L.}~\bibnamefont {D'Alessio}}, \bibinfo {author} {\bibfnamefont {Y.}~\bibnamefont {Kafri}}, \bibinfo {author} {\bibfnamefont {A.}~\bibnamefont {Polkovnikov}}, \ and\ \bibinfo {author} {\bibfnamefont {M.}~\bibnamefont {Rigol}},\ }\href {\doibase 10.1080/00018732.2016.1198134} {\bibfield  {journal} {\bibinfo  {journal} {Adv. Phys.}\ }\textbf {\bibinfo {volume} {65}},\ \bibinfo {pages} {239} (\bibinfo {year} {2016})},\ \Eprint {http://arxiv.org/abs/1509.06411} {arXiv:1509.06411 [cond-mat.stat-mech]} \BibitemShut {NoStop}%
\bibitem [{\citenamefont {Pastur}\ and\ \citenamefont {Mar\v{c}enko}(1967)}]{Pastur:1967zca}%
  \BibitemOpen
  \bibfield  {author} {\bibinfo {author} {\bibfnamefont {L.~A.}\ \bibnamefont {Pastur}}\ and\ \bibinfo {author} {\bibfnamefont {V.~A.}\ \bibnamefont {Mar\v{c}enko}},\ }\href {\doibase 10.1070/SM1967v001n04ABEH001994} {\bibfield  {journal} {\bibinfo  {journal} {Math. USSR Sb.}\ }\textbf {\bibinfo {volume} {1}},\ \bibinfo {pages} {457} (\bibinfo {year} {1967})}\BibitemShut {NoStop}%
\bibitem [{\citenamefont {Johnson}(2021{\natexlab{c}})}]{Johnson:2021zuo}%
  \BibitemOpen
  \bibfield  {author} {\bibinfo {author} {\bibfnamefont {C.~V.}\ \bibnamefont {Johnson}},\ }\href {\doibase 10.1103/PhysRevLett.127.181602} {\bibfield  {journal} {\bibinfo  {journal} {Phys. Rev. Lett.}\ }\textbf {\bibinfo {volume} {127}},\ \bibinfo {pages} {181602} (\bibinfo {year} {2021}{\natexlab{c}})},\ \Eprint {http://arxiv.org/abs/2106.09048} {arXiv:2106.09048 [hep-th]} \BibitemShut {NoStop}%
\bibitem [{\citenamefont {Johnson}(2021{\natexlab{d}})}]{Johnson:2021rsh}%
  \BibitemOpen
  \bibfield  {author} {\bibinfo {author} {\bibfnamefont {C.~V.}\ \bibnamefont {Johnson}},\ }\href@noop {} {\  (\bibinfo {year} {2021}{\natexlab{d}})},\ \Eprint {http://arxiv.org/abs/2104.02733} {arXiv:2104.02733 [hep-th]} \BibitemShut {NoStop}%
\bibitem [{\citenamefont {Saraswat}(2026)}]{Saraswat:2026cuo}%
  \BibitemOpen
  \bibfield  {author} {\bibinfo {author} {\bibfnamefont {K.}~\bibnamefont {Saraswat}},\ }\href@noop {} {\  (\bibinfo {year} {2026})},\ \Eprint {http://arxiv.org/abs/2601.20954} {arXiv:2601.20954 [hep-th]} \BibitemShut {NoStop}%
\bibitem [{\citenamefont {Johnson}\ \emph {et~al.}(2021)\citenamefont {Johnson}, \citenamefont {Rosso},\ and\ \citenamefont {Svesko}}]{Johnson:2021owr}%
  \BibitemOpen
  \bibfield  {author} {\bibinfo {author} {\bibfnamefont {C.~V.}\ \bibnamefont {Johnson}}, \bibinfo {author} {\bibfnamefont {F.}~\bibnamefont {Rosso}}, \ and\ \bibinfo {author} {\bibfnamefont {A.}~\bibnamefont {Svesko}},\ }\href {\doibase 10.1103/PhysRevD.104.086019} {\bibfield  {journal} {\bibinfo  {journal} {Phys. Rev. D}\ }\textbf {\bibinfo {volume} {104}},\ \bibinfo {pages} {086019} (\bibinfo {year} {2021})},\ \Eprint {http://arxiv.org/abs/2102.02227} {arXiv:2102.02227 [hep-th]} \BibitemShut {NoStop}%
\bibitem [{\citenamefont {Johnson}\ and\ \citenamefont {Rodrigues}(2026)}]{Johnson:2026jbq}%
  \BibitemOpen
  \bibfield  {author} {\bibinfo {author} {\bibfnamefont {C.~V.}\ \bibnamefont {Johnson}}\ and\ \bibinfo {author} {\bibfnamefont {J.}~\bibnamefont {Rodrigues}},\ }\href@noop {} {\  (\bibinfo {year} {2026})},\ \Eprint {http://arxiv.org/abs/2601.03351} {arXiv:2601.03351 [hep-th]} \BibitemShut {NoStop}%
\end{thebibliography}%


\end{document}